%% file: ANA-EXOT-2018-32-PAPER.tex
\pdfoutput=1
\newcommand*{\ATLASLATEXPATH}{}
\documentclass[cernpreprint, atlasdraft=false, texlive=2016, UKenglish]{\ATLASLATEXPATH atlasdoc}
\PreprintIdNumber{CERN-EP-2019-276}
 
\DeclareBoolOption[true]{LANGEDIT}

\usepackage{\ATLASLATEXPATH atlaspackage}
 
\makeatletter
\DeclareOldFontCommand{\rm}{\normalfont\rmfamily}{\mathrm}
\DeclareOldFontCommand{\sf}{\normalfont\sffamily}{\mathsf}
\DeclareOldFontCommand{\tt}{\normalfont\ttfamily}{\mathtt}
\DeclareOldFontCommand{\bf}{\normalfont\bfseries}{\mathbf}
\DeclareOldFontCommand{\it}{\normalfont\itshape}{\mathit}
\DeclareOldFontCommand{\sl}{\normalfont\slshape}{\@nomath\sl}
\DeclareOldFontCommand{\sc}{\normalfont\scshape}{\@nomath\sc}
\makeatother
 
\usepackage{\ATLASLATEXPATH atlasbiblatex}
 
\usepackage{\ATLASLATEXPATH atlasphysics}
 
\graphicspath{{logos/}{figures/}}

\newcommand{\bd}{\begin{description}}
\newcommand{\ed}{\end{description}}
\newcommand{\bi}{\begin{itemize}}
\newcommand{\ei}{\end{itemize}}
\newcommand{\beq}{\begin{equation}}

\chardef\til=126
 
\newcommand{\mjj}{{m_{\rm jj}}}
\newcommand{\mX}{{m_{\rm X}}}
\newcommand{\sigmaX}{{\sigma_{\rm X}}}
\newcommand{\lumi}{139}

\AtlasTitle{
Search for dijet resonances in events with an isolated charged
lepton using $\sqrt{s} = 13$~TeV proton--proton collision  data collected by the ATLAS detector
}
 
\AtlasVersion{4.0}
\AtlasAbstract{
A search for dijet resonances in events with at least one isolated charged
lepton is performed using $139~{\text{fb}}^{-1}$ of $\sqrt{s}=13$~TeV
proton--proton collision data recorded by the ATLAS detector at the LHC.
The dijet invariant-mass ($\mjj$) distribution constructed from events with at
least one isolated electron or muon is searched in the region
$0.22 < \mjj < 6.3$~TeV for excesses above a smoothly falling background from
Standard Model processes.
Triggering based on the presence of a lepton in the event reduces limitations
imposed by minimum transverse momentum thresholds for triggering on jets.
This approach allows smaller dijet invariant masses to be probed than in
inclusive dijet searches, targeting a variety of new-physics models, for
example ones in which a new state is produced in association with a
leptonically decaying $W$ or $Z$ boson.
No statistically significant deviation from the Standard Model background
hypothesis is found.
Limits on contributions from generic Gaussian signals with widths ranging from
that determined by the detector resolution up to 15\% of the resonance mass
are obtained for dijet invariant masses ranging from 0.25~TeV to 6~TeV.
Limits are set also in the context of several scenarios beyond the Standard
Model, such as the Sequential Standard Model, a technicolor model, a
charged Higgs boson model and a simplified Dark Matter model.
}
 
\author{The ATLAS Collaboration}
 
\AtlasRefCode{EXOT-2018-32}

\AtlasJournalRef{JHEP 06 (2020) 151}
 
\AtlasDOI{10.1007/JHEP06(2020)151}
 
\PreprintIdNumber{CERN-EP-2019-276}

\AtlasCoverCommentsDeadline{XX January 2020}
 
\AtlasCoverSupportingNote{ATL-COM-PHYS-2018-1524}{https://cds.cern.ch/record/2646242}

\AtlasJournal{JHEP}
 
\AtlasCoverAnalysisTeam{Vincent Pascuzzi, Sergei Chekanov, Dylan Frizzell, Jeremy Love, Adriana Milic, Wasikul Islam, Alexander Khanov, Peter Krieger, Henry Meng, James Proudfoot, Rui Wang, Julian Glatzer, Lluisa Maria Mir Martinez, Jana Schaarschmidt, Max Isacson}
 
\AtlasCoverEdBoardMember{Dmitri Tsybyshev (chair)}
\AtlasCoverEdBoardMember{Francesco Guescini}
\AtlasCoverEdBoardMember{Lailin Xu}

\AtlasCoverEgroupEditors{atlas-EXOT-2018-32-editors@cern.ch}
 
\AtlasCoverEgroupEdBoard{atlas-EXOT-2018-32-editorial-board@cern.ch}

\hypersetup{pdftitle={ATLAS document},pdfauthor={The ATLAS Collaboration}}
 
\begin{document}
 
\maketitle

\section{Introduction}
 
In the Standard Model (SM),
events with two or more jets are usually produced by strong interactions
described by quantum chromodynamics (QCD).
Searches for resonances in dijet invariant-mass distributions provide a means to investigate a wide
range of theories beyond the Standard Model (BSM).
Such searches are sensitive to heavy particles
that decay into two partons which, following fragmentation, form two jets.
Studies of this kind were among
the first published using early data from the ATLAS~\cite{PhysRevLett.105.161801, Aad:2011fq, Aad:2011aj,ATLAS:2012pu}
and CMS~\cite{Khachatryan:2010jd,Chatrchyan:2011ns,CMS:2012yf} experiments at the Large Hadron Collider
(LHC) at CERN, when operations at high energy first began.
Later results used new datasets as the LHC increased the collision energy, from 7~TeV to 8~TeV
during Run~1~\cite{Aad:2014aqa, Sirunyan:2018pas, Khachatryan:2016ecr},
and then to 13~TeV for Run~2~\cite{highmass,Aaboud:2017yvp,Aaboud:2018fzt,2017520,Sirunyan:2018xlo,Aad:2019hjw,Sirunyan:2019vgj,CMS-EXO-19-004}.
Typically, these searches initially focused on resonances with high dijet invariant masses, $\mjj$,
e.g. on enhanced event yields in the new kinematic regime opened up by the increase in energy.
However, as the integrated
luminosity collected at the highest available energy increased, without signs of new physics
at high masses, there has been renewed interest in exploiting these large datasets
to also look for signals in the region $\mjj < 1$~TeV.
These low-mass regions are difficult to access in inclusive dijet searches due to limitations imposed by the
high transverse-momentum thresholds applied to jet triggers to keep trigger rates at
manageable levels.
Such studies have been performed with inclusive samples using analyses done at so-called `trigger
level'~\cite{Aaboud:2018fzt,Khachatryan:2016ecr}.
Other strategies involve requiring the
presence of an associated object that can be used for triggering, such as a photon
~\cite{Aaboud:2019zxd,Sirunyan:2019sgo} or a jet \cite{Sirunyan:2017dnz,Sirunyan:2017nvi,Sirunyan:2018ikr,Sirunyan:2019vxa}
from initial-state radiation.
The results presented here complement these techniques by searching for dijet resonances in events containing
an isolated electron ($e$) or muon ($\mu$).
Besides providing access to lower dijet invariant masses, the requirement of a final-state lepton
in addition to jets provides sensitivity to a set of new-physics models that cannot be studied using jet triggers.

\begin{figure}
\begin{center}
 
\subfloat[] {
\includegraphics[width=0.4\textwidth]{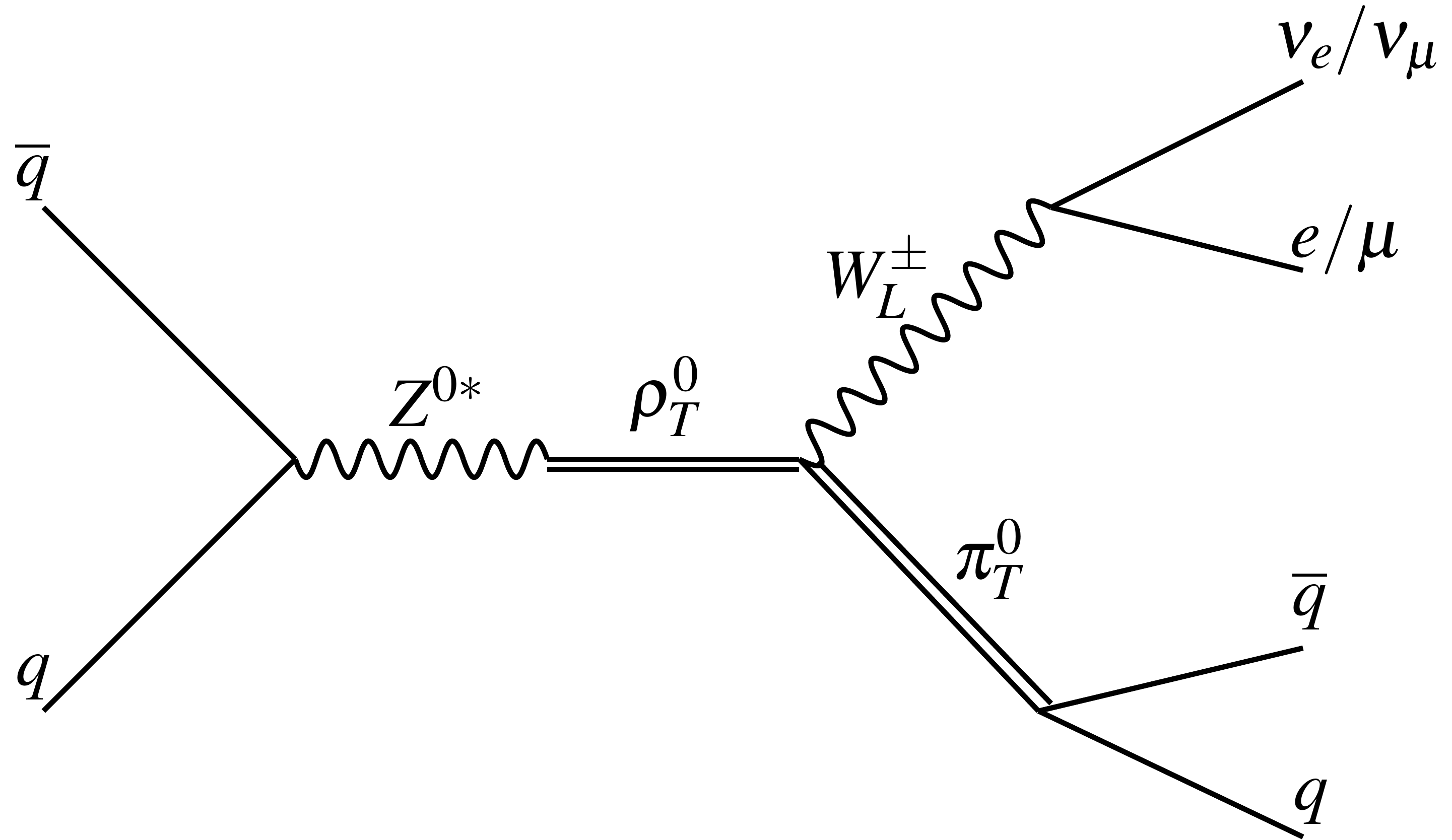}\hfill
}
\subfloat[] {
\includegraphics[width=0.4\textwidth]{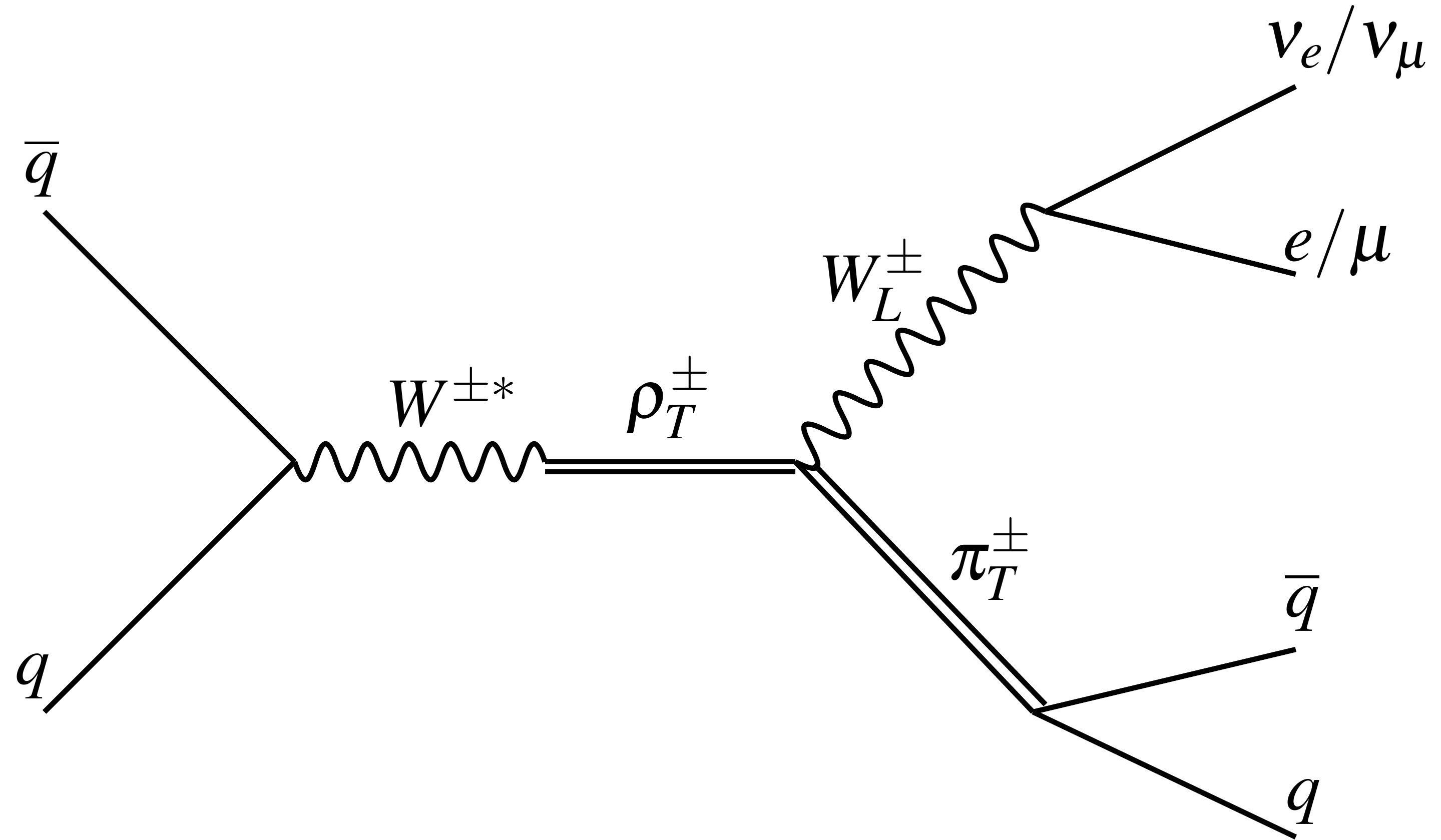}\hfill
}

\subfloat[] {
\includegraphics[width=0.4\textwidth]{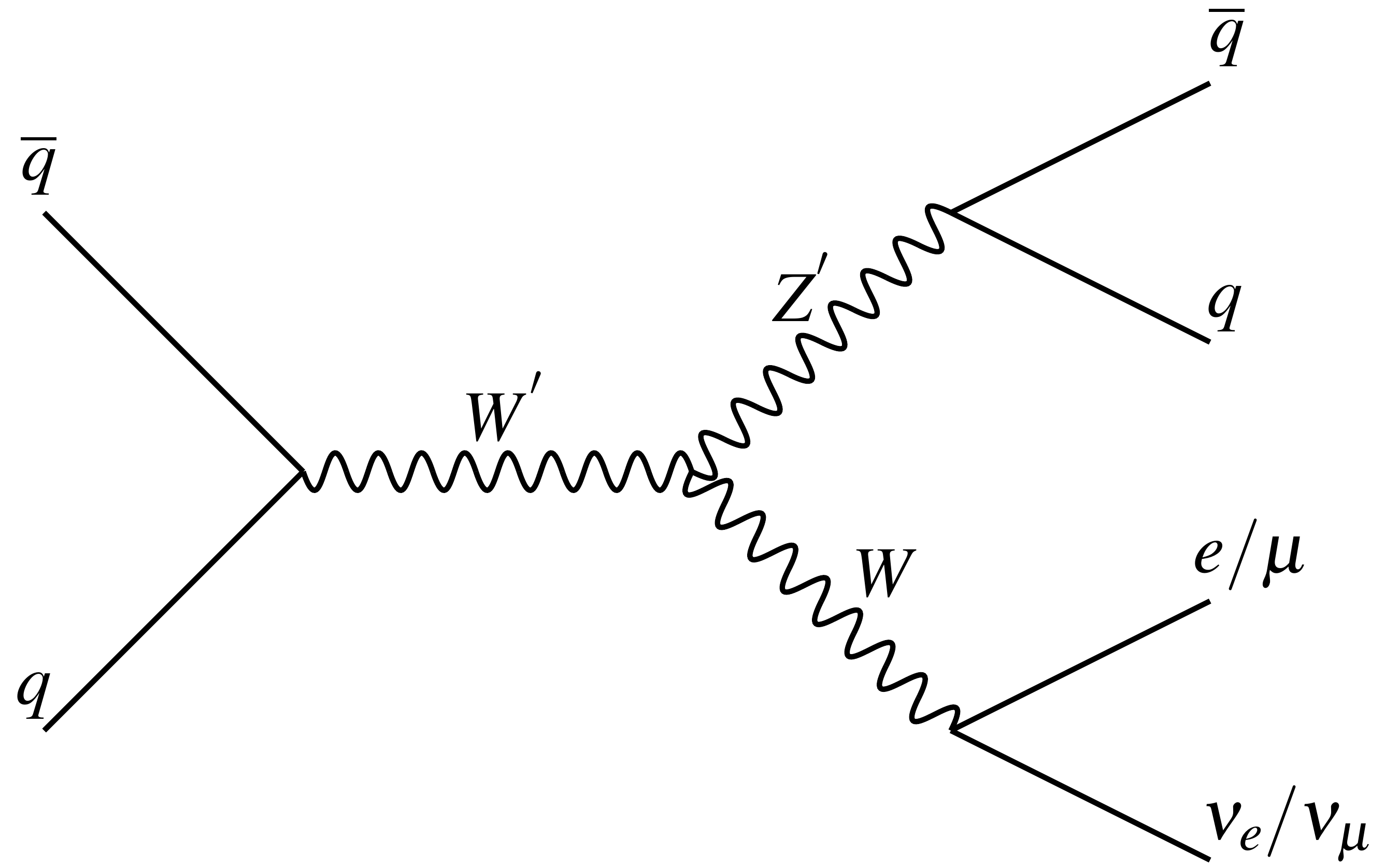}\hfill
}
 
\subfloat[] {
\includegraphics[width=0.4\textwidth]{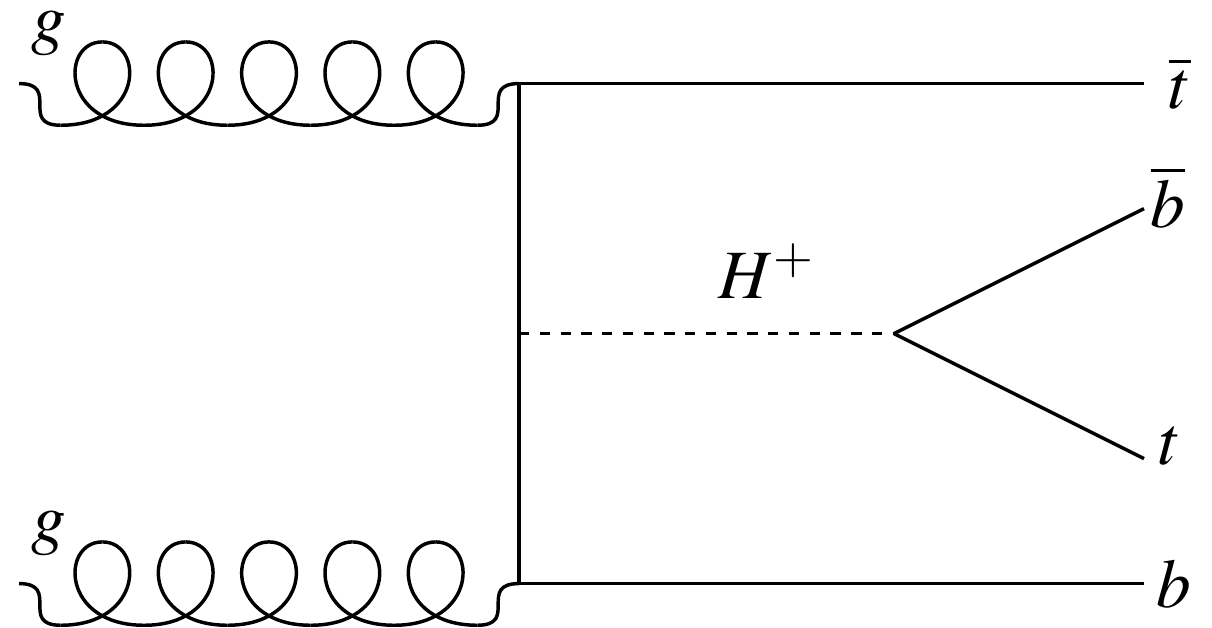}\hfill
}
 
\subfloat[] {
\includegraphics[width=0.4\textwidth]{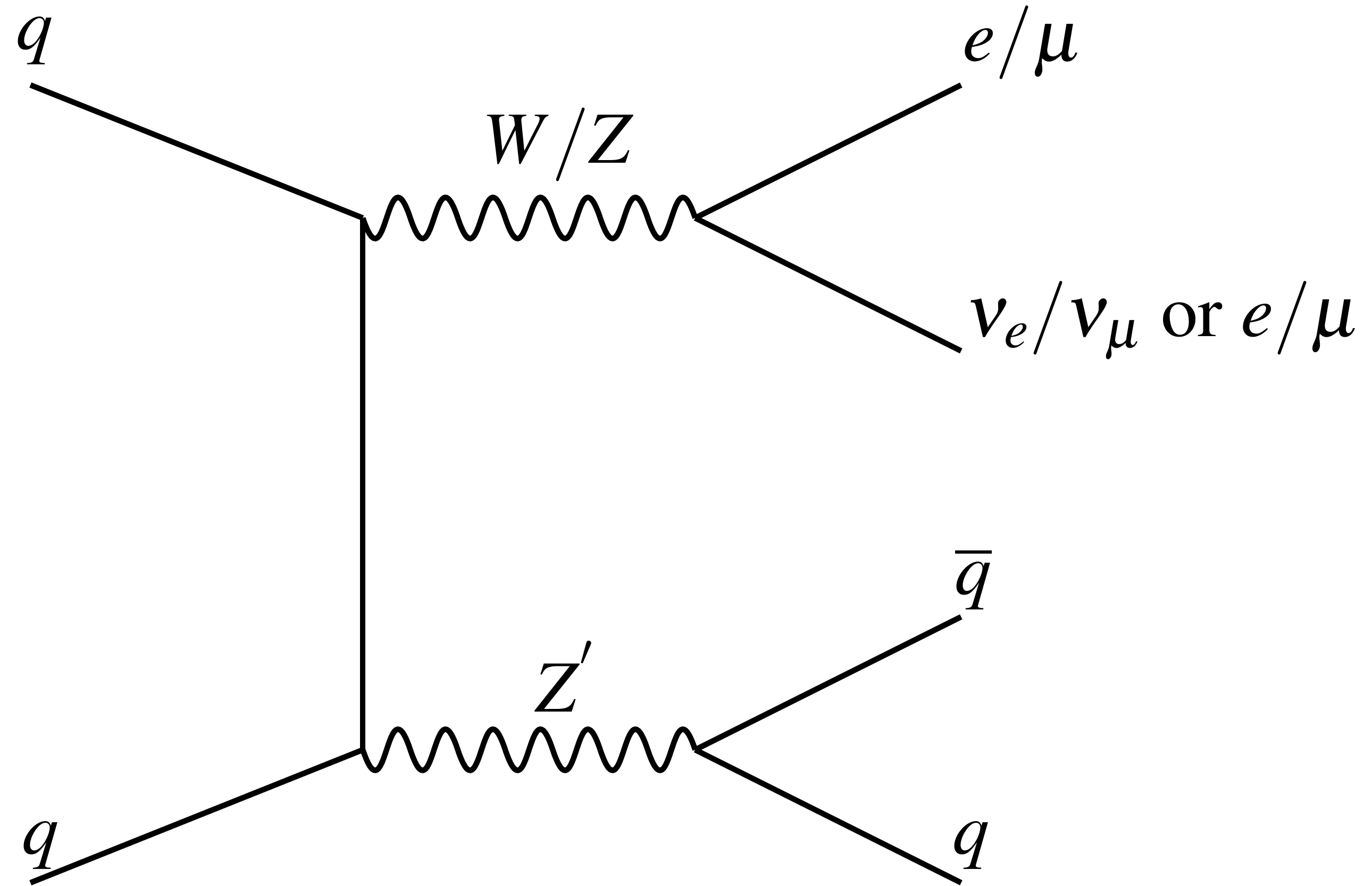}\hfill
}
 
\end{center}
\caption{
Representative Feynman diagrams for the processes considered in this analysis: (a)-(b) the techicolor model with
production of $\rho_T$ decaying into $\pi_T W^{\pm}$, (c)
$W' \to Z' W^{\pm}$ production
in the Sequential Standard Model, (d) the charged Higgs boson production in association with a top quark, $tbH^{+}$,
(e) the simplified dark-matter model.
}
\label{fig:feynman}
\end{figure}

Many BSM models predict new heavy resonances in production modes, yielding a final state consisting of
jets, produced in the resonance decay, accompanied by at least one lepton. At hadron colliders, possible
processes are
$q\bar{q}'\to W X\to \ell\nu q\bar{q}$ and $q\bar{q}'\to X' \to W X \to \ell\nu q\bar{q}$,
as well as production induced by gluon--gluon fusion, where the
$X$ and $X'$ can be either scalar or vector particles.
Examples include technicolor
models, $q\bar{q} \to \rho_{T} \to W\pi_{T} \to \ell\nu q\bar{q}$~\cite{Eichten:1997yq} (in which a
technirho $\rho_{T}$ decays into a $W$ boson and a technipion $\pi_{T}$, see Figure~\ref{fig:feynman}(a-b)),
the Sequential Standard Model~\cite{AltarelliMele},
$W'\to W Z' \to l\nu q\bar{q}$ (where $W'$ and $Z'$ are new heavy gauge
bosons, see Figure~\ref{fig:feynman}(c)), and charged Higgs models~\cite{Akeroyd:2016ymd} (Figure~\ref{fig:feynman}(d)).
A number of dark-matter (DM) models also predict new resonances that can be produced
in association with vector bosons~\cite{Abercrombie:2015wmb} (Figure~\ref{fig:feynman}(e)).

Searches for the signatures of specific  models may benefit by requiring  the presence of
other leptons (as in the case of associated $Z$ boson production), $b$-tagged jets,
missing transverse momentum (in the case of $W$ boson or top decays) or other model-specific kinematic
quantities.
However, the study presented in this article focuses on a generic search for BSM resonances in the dijet
invariant-mass distribution constructed from events with at least one isolated electron or muon, in
order to explore the potential of searches without signal-specific selections.
Model-dependent limits are
set by taking into account the signal shapes expected in the
models described above.
This search uses an integrated luminosity of $\lumi~\text{fb}^{-1}$ of $\sqrt{s}=13$~TeV proton--proton
($pp$) collision data recorded by the ATLAS detector over the full period of Run 2 of the LHC.

\section{ATLAS detector}
The ATLAS detector~\cite{PERF-2007-01,ATLAS-TDR-2010-19,PIX-2018-001} consists of an inner tracking detector
(ID), surrounded by
a superconducting solenoid that provides a 2\,T magnetic field, electromagnetic and hadronic
calorimeters, and a muon spectrometer (MS).
The ID provides tracking in the
pseudorapidity\footnote{ATLAS uses a right-handed coordinate system with its origin at the nominal
interaction point (IP) in the centre of the detector and the $z$-axis along the beam line.
The
$x$-axis points from the IP to the centre of the LHC ring, and the $y$-axis points upwards.
Cylindrical coordinates $(r,\phi)$ are used in the transverse plane, $\phi$ being the azimuthal
angle around the beam line.
Transverse momentum and energy are defined as $p_\text{T}=p\sin
\theta$ and $E_\text{T}=E \sin \theta$, respectively. The pseudorapidity is defined in terms of
the polar angle $\theta$ as $\eta = -\ln \tan(\theta/2)$. The angular separation between
two objects in $\eta$--$\phi$ space is defined as $\Delta R = \sqrt{(\Delta \eta)^2 + (\Delta\phi)^2}$.}
region $|\eta| \le 2.5$ and consists of
silicon pixel and microstrip detectors surrounded by a transition radiation tracker, which also provides
information for electron identification.
Each tracking detector consists of a central barrel and two endcap sections.
 
The electromagnetic calorimeter is a sampling device made of lead absorbers with liquid argon
(LAr) as active medium.
It comprises a barrel ($|\eta| \le 1.475$) and two endcaps ($1.375 \le |\eta| \le 3.2$).
To facilitate corrections for energy losses upstream of the
calorimeter, the cryostat is equipped with a presampler layer in the region $|\eta| \le 1.8$.
Hadronic sampling calorimetry is provided by a steel and scintillator-tile calorimeter in the
region $|\eta| \le 1.7$, complemented by a copper/LAr system in the region $1.5 \le|\eta|\le3.2$.
The forward region ($3.1 \le |\eta| \le 4.9$) is equipped with both electromagnetic and hadronic
calorimeters composed of copper/LAr and tungsten/LAr, respectively.
 
The muon spectrometer is the outermost ATLAS subsystem.
It detects muons in the pseudorapidity region up to $|\eta| = 2.7$, with triggering capability up to $|\eta| = 2.4$.
The MS consists of a barrel ($|\eta| \le 1.05$) and two endcap
sections ($1.05 \le |\eta| \le 2.7$). A system of three large superconducting air-core toroid magnets,
each with eight coils, provides a magnetic field with a bending integral of about 2.5~Tm in
the barrel and up to 6~Tm in the endcaps.
 
The trigger system~\cite{Aaboud:2016leb} consists of a first-level trigger implemented in hardware
using a subset of the detector information to
accept events from the 40~MHz bunch crossings at a rate of 100~kHz,
followed by a
software-based trigger implemented in a large computer farm,
which reduces the acceptance rate so that events are recorded at about 1~kHz.
 
\section{Object definitions and event selection}
 
The analysis presented in this paper is based on data collected with the ATLAS detector during the
2015--2018 data-taking period, referred to as Run 2.
The data were recorded under stable beam
conditions while all relevant subdetectors were fully operational, and were subject to detailed
quality checks.
The data sample corresponds to an integrated luminosity of $\lumi$~fb$^{-1}$
with an uncertainty of 1.7\%.
This uncertainty was derived from calibration of the luminosity scale
using $x$--$y$ beam-separation scans, following a methodology similar to that detailed in
Ref.~\cite{DAPR-2013-01} using data from the LUCID-2 detector~\cite{LLUCID2} for the baseline
measurement.
 
Candidate events were accepted by either single-muon or single-electron triggers~\cite{Aaboud:2016leb}
with various transverse momentum $p_\text{T}$ (muons) or transverse energy $E_\text{T}$ (electrons)
thresholds, as well as data quality and lepton isolation requirements.
The lowest $p_\text{T}$ ($E_\text{T}$)
threshold without trigger prescaling is 24 (26)~GeV and includes a lepton isolation
requirement that is not applied for triggers with higher thresholds.
A trigger matching requirement \cite{Aaboud:2016leb} is applied where
the reconstructed lepton must lie within the vicinity of the corresponding trigger-level object.
This requirement reduces the fake rate and allows a consistent definition of efficiencies between Monte Carlo simulations and data.
 
Muons are reconstructed by using a combined fit to hits in the ID and MS,
and are required to have $p_\text{T}\ge 7$~GeV and $|\eta|\le 2.5$ and to satisfy `medium' quality criteria, in order to select particles from the primary interaction~\cite{Aad:2016jkr}.
Identification requirements based on the number of hits
in the ID and MS subsystems, as well as the significance of the difference
$|q/p_\text{MS} - q/p_\text{ID}|$, where $q$ is the charge and $p_\text{MS}$ ($p_\text{ID}$)
is the momentum as measured in the MS (ID), are applied to the combined tracks.
Muon tracks are required to satisfy $|d_0/\sigma(d_{0})|\le 3$ and $|z_0 \times \sin\theta|\le 0.5$~mm,
where $d_0$ is the transverse impact parameter relative to the beam line, $\sigma(d_0)$
is its uncertainty, $z_0$ is the distance along the beam line to the primary vertex
from the point where $d_0$ is measured, and $\theta$ is the polar angle.
The primary vertex is chosen as the vertex with the highest $\sum{\pt^2}$, where
the sum is over tracks associated with that vertex and having $\pt>500$~MeV; at least two such tracks are required.
 
Electrons are identified as energy clusters formed in the electromagnetic calorimeter~\cite{Aad:2019tso} 
matched to tracks in the ID, with requirements of $E_\text{T} > 7$~GeV and $|\eta|\le 2.47$.
Candidate electrons must have tracks satisfying $|d_0/\sigma(d_{0})| \le 5$ and
$|z_0 \times \sin\theta| \le 0.5$~mm and meet `tight' quality criteria~\cite{Aad:2019tso}.  
Electrons within the barrel--endcap transition region of the electromagnetic calorimeter, $1.37 \le |\eta| \le 1.52$,
or which share a track with an identified muon, are discarded.
 
Leptons are further required to be isolated from other objects in the event using
$\pt$-dependent criteria based on calorimeter and tracking information.
The isolation parameters were  tuned to provide a constant efficiency as a function of transverse momentum,
and the highest background rejection below 60~GeV (`FCTight'). 
The lepton isolation and $\pt$ requirements allow a consistent definition of lepton candidates when considering
data-taking periods in which different trigger configurations were used.
Lepton misidentification rates for the chosen isolation requirements are discussed in Refs.~\cite{Aad:2016jkr,Aad:2019tso}.
The combined trigger efficiency is about 88\%, averaged over the two lepton flavours.

Jets are reconstructed using the anti-$k_t$ algorithm~\cite{Cacciari:2008gp} with a radius parameter of
$R=0.4$, as implemented in the FastJet package~\cite{fastjet}, using topological clusters of energy deposits
in the calorimeters~\cite{Aad:2016upy} as inputs.
These jets are corrected for contributions arising from additional collisions in the same and nearby bunch
crossings (pile-up)~\cite{Aad:2015ina}, and calibrated to the particle energy scale (i.e. before interaction with the detector)~\cite{Aaboud:2017jcu}.
Jet candidates are required to have $p_\text{T}\ge 20$~GeV and to be within $|\eta|\le 2.4$.
To suppress jets arising from pile-up, a jet-vertex-tagging technique~\cite{ATLAS-CONF-2014-018} is
applied to jets with $p_\text{T}\le 60$~GeV, requiring that at least 60\% of the total $\pt$ of tracks
in the jet be associated with the primary vertex in each event.
Candidate jets with fewer than three
associated tracks are discarded if they lie within a cone of $\Delta R=0.2$ around a muon candidate,
and disregarded irrespective of the track requirement for the electron candidates.
The track-multiplicity requirement is effective in rejection of  misreconstructed jets originating from muons.
Electron and muon candidates are discarded if they are within a cone of $\Delta R=0.4$ around a remaining jet's axis.
 
Following this object selection, a signal region is defined by requiring at least one isolated lepton ($e$ or $\mu$)
with $p_\text{T}^\ell \ge 60$~GeV and at least two jets.
Dijet invariant masses are constructed by combining the two jets having the highest $p_\text{T}$.
Only events with $\mjj \ge 0.22$~TeV are considered;
this minimum value is chosen in order to avoid
the low-$\mjj$ region where the event rate does not
monotonically decrease due to a kinematic bias in jet $\pt$ from the minimum $p_\text{T}^\ell$ of leptons. The $\mjj$ mass distribution that includes a
region below $0.22$~TeV
is difficult to describe with a function that falls smoothly over the entire range of $\mjj$.
 
\section{Monte Carlo simulations}
\label{sec:mc}
 
Monte Carlo (MC) simulations are used to investigate
contributions to the $\mjj$ distribution from various SM processes, as well as
to estimate the background from data.
The sources of background modelled using MC simulation are the QCD multijet, $t\bar{t}$ and $W$+jets processes.
The multijet event sample was simulated with the {\sc Pythia 8.186}~\cite{Sjostrand:2007gs}
generator with the NNPDF2.3~\cite{Ball:2012cx} set of parton distribution functions (PDF) and a set of tuned
parameters called the A14 tune~\cite{ATL-PHYS-PUB-2014-021}.
The $t\bar{t}$ and $W$+jet events were produced using the {\sc PowhegBox}~\cite{Nason:2004rx,Frixione:2007vw,Alioli:2010xd,Frixione:2007nw} v2 generator
interfaced with {\sc Pythia 8}.
This simulation used the CT10 NLO PDF set~\cite{Lai:2010vv} and the AZNLO tune~\cite{Aad:2014xaa}.
 
For the model-dependent searches, MC simulations are also used to predict the expected signal shapes
for the BSM models discussed earlier:
(1) $W'\to W Z' \to \ell\nu q\bar{q}$;
(2) $\rho_{T} \rightarrow W^{\pm} \pi_{T} \to \ell\nu q\bar{q}$;
(3) charged Higgs boson production in association with a top quark, $tbH^{+}$; and
(4) a simplified DM model with an axial-vector mediator, $Z'$.
 
The $W'\to W Z' \to \ell\nu q\bar{q}$ and $\rho_{T} \rightarrow W^{\pm} \pi_{T} \to \ell\nu q\bar{q}$
simulations are performed at leading-order QCD using {\sc Pythia 8} and {\sc Pythia 6}~\cite{pythia6}, respectively, with
the CT10 NLO PDF set.
The parton showering and hadronisation of the latter simulation are performed using the A14 tune of {\sc Pythia 8}.
The first model assumes a $Z'$ dijet resonance produced in association with
a leptonically decaying $W$ from the $q\bar{q}\to W'$ process.
The relative width of the $Z'$ is set to 3.2\%, which is the default value in {\sc Pythia}.
The $W'$ to $W Z'$ branching fraction is chosen to be 0.5 and the mass difference between the
$W'$ and $Z'$ was set to 250~GeV.
This latter requirement yields the largest predicted cross-section
for the desired final state.
This model is also used to estimate systematic uncertainties in generic signals approximated by Gaussian functions with widths varying between 5\% and 15\% of the dijet invariant mass.
 
The second model considered is a generic technicolor model~\cite{Lane:2016kvg} that assumes the production of a technirho,
$\rho_{T}$, that decays into a leptonically decaying $W$ boson and a technipion $\pi_{T}$,
decaying into two jets.
The mass of the $ \rho_{T}$ is chosen to be a factor of two larger than the mass of the $\pi_{T}$,
which maximises the cross-section for the $l\nu q\bar{q}$ final state.
The signal width for this model is approximately 15\% of the predicted technipion mass.
 
The $tbH^{+}$ process is modelled with \MGMCatNLO~\cite{Alwall:2014hca} at next-to-leading
order (NLO) in QCD~\cite{Degrande:2015vpa}, based on a two-Higgs-doublet model (2HDM) in
the $m_h^{\mathrm{mod-}}$ scenario \cite{Carena:2013ytb}
and a four-flavour scheme
implementation with the NNPDF2.3 PDF set.
The $H^+$ decay into $t\bar{b}$ is assumed.
Parton showering and hadronisation are modelled using
{\sc Pythia 8} with the A14 tune.
This simulation uses the narrow-width approximation~\cite{Uhlemann:2008pm},
and effects related to $W$ boson polarisation and the interference between
$tbH^{+}$ and
the SM $ t\bar{t} + b\bar{b} $ background are not included.
The narrow-width approximation has a  negligible impact on the limit presented in this paper
since the peak in the $\mjj$ distribution has a relative half width
at half maximum (HWHM) of
about $30\%$ of the $\mjj$ peak position.
This width  is much larger than the $H^{+}$ natural width of about $4\%$ of the $H^{+}$ mass for $\tan \beta=1$, where $\tan \beta$
is the ratio of the vacuum expectation values of the two scalar doublets in the 2HDM.
While the $t\bar{b}$ final state includes more than two jets, simulation studies indicate that, for
events arising from the $tbH^+$ process, $\mjj$ distributions obtained from the two highest-$\pt$
jets have well-defined peaks for $m_{H^+}$ values above 600~GeV.
The reconstructed $\mjj$ peak position is shifted from the $m_{H^+}$ value
by about 30\% to lower masses
since the two leading jets used for $\mjj$ do not contain the complete information about the $H^+$ decays.

This analysis also studies a benchmark simplified DM model with an axial-vector mediator $Z'$, in which
the lepton originates from the decay of a $W$ boson, e.g.
$q\bar{q}\rightarrow Z'W$ where the $Z'$ decays into jets and the $W$ via $W\rightarrow \ell\nu$.
This model assumes the leptophobic couplings $g_q = 0.25$, $g_\ell = 0$, and $g_\text{DM} = 1$, following the recommendations
of the LHC-DMWG~\cite{Abercrombie:2015wmb}.
Here $g_q$, $g_\ell$ and $g_\text{DM}$ are the couplings of the mediator to quarks, leptons and the DM particle, respectively.
Signal MC events were generated with \MGMCatNLO at leading order in QCD.
Possible interference between SM and DM processes was studied at parton level and shown not to affect the shape or
normalisation of the resonance mass peaks, so is not included in the simulation sample.
 
All MC samples, with the exception of the $tbH^{+}$ and  DM models, were passed through the full ATLAS detector
simulation~\cite{Aad:2010ah} based on {\sc Geant4}~\cite{Agostinelli:2002hh};
the $tbH^{+}$ and DM processes were simulated using the ATLAS fast simulation framework,
ATLFAST-II~\cite{SOFT-2010-01}, which uses parameterisations of electromagnetic and hadronic
showers in the calorimeters.
All simulated events are corrected so that the object identification, reconstruction and trigger efficiencies, energy scales and energy resolutions match those determined from data control samples.
After this correction, additional systematic uncertainties were applied to cover for the difference between the data and MC
simulations as described in Section~\ref{sec:sys}.
Additional simulated $pp$ collisions generated using {\sc Pythia 8},
with the A3 set of tuned parameters~\cite{ATL-PHYS-PUB-2016-017} and the NNPDF2.3 PDF set,
were overlaid to simulate the effects of pile-up in a manner that matches the multiplicity distribution of additional collisions
in the data.
Simulated events were reconstructed and analysed with the same algorithms as used for data.
 
\section{Analysis procedure}
\label{sec:ana}
 
The existence of a new resonant state of mass $\mX$ decaying into partons that hadronise to two jets
could lead to an observable excess of events at $\mjj\approx\mX$ on an otherwise smooth and monotonically
decreasing dijet invariant-mass distribution.
This analysis presents a search for such an excess in the range $0.22 < \mjj < 6.3$~TeV.
 
The bin widths of the $\mjj$ distribution are chosen to be approximately equal to the dijet mass resolution at
a given mass and therefore widen from 13~GeV to 120~GeV, over the specified range in $\mjj$.
The following fit function~\cite{Aad:2014aqa,Aaboud:2018fzt,Aaboud:2019zxd} is used to model the shape of the
estimated background,
\begin{equation}
f(x) = p_1 (1 - x)^{p_2} x^{p_3 + p_4\ln x + p_5 \ln^2 x},
\label{eq:function}
\end{equation}
where $x \equiv \mjj /\sqrt{s}$ and the $p_i$\ are free parameters.
 
To investigate the ability of Eq.~(\ref{eq:function}) to accurately describe the
background in the signal region, a likelihood
fit is performed to the background estimate obtained from MC simulations of QCD multijet, $W+$jets and
$t\bar{t}$ samples.
The combined contribution from $W$ and top-quark processes in the MC simulations
varies from 1\% to 10\% as a function of $\mjj$.
The five-parameter fit function provides a good description of the signal region in this MC sample.
However, Eq.~(\ref{eq:function}) with $p_5=0$ also provides an adequate description of this distribution, which has
far fewer events than are available in data.
Additional studies were therefore undertaken, as described below.
One is based on a control region (CR) defined for MC simulation, and the other on a CR defined for data;
both provide far more events than are available in the MC distribution discussed above.
These studies are also used to investigate possible
effects related to jet reconstruction, or to leptons misidentified as jets, which may lead to
structures in the $\mjj$ distribution that can be difficult to describe with a smoothly falling
distribution and could be misinterpreted as a potential signal.
In the first of these, a three-jet control region, referred to as the `2+1 jets' CR is constructed using MC events.
This is identical to the signal region with the exception that a third jet with $p_\text{T} > 60$~GeV is
required instead of the charged lepton.
Removing the requirement of a final-state lepton increases the number of events in the dominant multijet background sample by
more than an order of magnitude.
No significant deviations of the MC distribution from the fit hypothesis of Eq.~(\ref{eq:function}) are observed in
the 2+1 jets CR.
The composition of this CR illustrates
the decreased relative contributions from $W$ and top-quark processes when requiring a third jet instead of a final-state lepton.
Here, in the 2+1 jets CR,  $W$ and top processes contribute to the overall background at a level of less than 0.1\% relative to QCD multijets.
 
The fit hypothesis of Eq.~(\ref{eq:function}) models the 2+1 jets CR well, giving
$\chi^2/\text{ndf}\simeq 1.2$ (where $\chi^2=130$ and $\text{ndf}=109$)
and a ratio of the MC distribution to the fit that is within 5\% of unity over the full range of $\mjj$.
The distribution of the fit residuals is consistent with  a  normal distribution with a mean of zero.
It is observed that all five
fit parameters are strongly constrained by the low-mass region, $\mjj < 1$~TeV, which has the most events.
Functions of the form of Eq.~(\ref{eq:function})  with fewer than five parameters
fail to adequately describe the dijet mass distribution in this CR.
 
To complement the 2+1 jets CR used in MC studies, a `loose electron' control region (LE-CR) is defined
for the data.
It is populated by selecting dijet events with at least one electron that satisfies
a set of loose identification criteria but not the more stringent tight identification criteria, ensuring
orthogonality with the signal region.
The LE-CR is expected to have an increased multijet
contribution due to the contamination from misidentified electrons.
The main goal of this region is to verify that the data do not show structures that can be interpreted as signals when using
the analytic fits.
The number of events in the
LE-CR is a factor of ten larger than in the
2+1 jets CR used in MC studies.
Studies of this control region demonstrate
that jet reconstruction does not lead to
structures that can be interpreted as signals.
The fit hypothesis of Eq.~(\ref{eq:function}) models the LE-CR
with $\chi^2/\text{ndf}\simeq 1.6$, where
$\chi^2=172$ and $\text{ndf}=109$ (see the Appendix~\ref{appendix:le}). The residuals are distributed according to a normal distribution with a mean
consistent with zero.
The fit with Eq.~(\ref{eq:function}) performs poorly when the minimum value of $\mjj$ is below 200~GeV.
The minimum value of $\mjj$ value is therefore set to 216 GeV, which is defined by the lower bin
edge of the chosen binning.
Fit functions with $p_5=0$ fail to describe the LE-CR.
 
Several alternative five-parameter functions for the description of the LE-CR region
have been investigated. It was found that the only five-parameter function that adequately describes the LE-CR as a whole,
and shows some systematic difference with respect to Eq.~(\ref{eq:function}) in the tail of the $\mjj$ distribution, is a function
obtained after replacing $p_5\ln^{2}x$ to $p_5/\sqrt{x}$.
The application of this alternative function to the LE-CR region
leads to residuals distributed according to a normal distribution with
a mean consistent with zero.
Functions with more than five parameters have also been investigated.
A test based on the Wilks' theorem \cite{wilks} shows that no additional parameters are needed.

To investigate potential biases in the description of the LE-CR by
Eq.~(\ref{eq:function}), `signal-injection' and `closure' tests are performed.
For the signal-injection test, signal events modelled according to Gaussian
distributions are added to the expected background distribution to assess
whether or not the correct numbers of events can be extracted using
signal-plus-background fits, assuming the known Gaussian signal shape.
For the closure test, signal-plus-background fits are run on the background-only
spectra of the LE-CR for different signal masses and the extracted signal yield
is taken as an estimate of a false signal.
In the first case, the extracted signal yield is consistent with the injected
number of events within the statistical uncertainty.
The number of extracted events in the closure tests was significantly
smaller than the statistical uncertainty of the data points.
This event rate for the signal region is considered as a source of
systematic uncertainty in the limit values (discussed in Section~\ref{sec:sys}).
 
Based on the studies of the control regions, the background-only hypothesis for the signal region is
constructed using Eq.~(\ref{eq:function}), over the $\mjj$ range from 216~GeV to 6.3~TeV.
 
To determine if the data deviate significantly from the
background-only hypothesis prediction, the
BumpHunter~\cite{bumphunter} test is used.
This test calculates the significance of any excess found in  mass
intervals in all possible locations of the binned $\mjj$.
The width of the search window varies from a minimum of two $\mjj$
bins up to half the extent of the full $\mjj$
mass distribution.
For each of the chosen intervals in $\mjj$, a local $p$-value is calculated from a
unique hypothesis test statistic.
The method takes into account the look-elsewhere effect~\cite{Gross:2010qma} by combining each of the
hypothesis tests to form a new hypothesis test, and calculating the minimum $p$-value amongst all tests.
A global $p$-value is then calculated and transformed to a significance assuming that bin-by-bin
fluctuations of the data follow a Poisson distribution.
Pseudo-experiments are then used to determine the most significant local excess and, finally, a global
significance is calculated.
A null result from the BumpHunter test can only be used as an indication for non-observation of the searched-for
processes but not as the necessary condition.
Signal-injection tests using the BumpHunter background-only fit show a lower signal extraction efficiency
for wide signals than for the signal-plus-background fit.

The Bayesian Analysis Toolkit \cite{Caldwell:2008fw} is used to set 95\%
credibility level (CL) upper limits on the cross-section for new processes that
have a signature of a new particle decaying into partons which fragment to two
jets in events with at least one isolated lepton of $p_\text{T}^\ell>60$~GeV.
In the case where no statistically significant deviations are observed according to the BumpHunter test, the exclusion limits are set at the 95\% CL on the production cross-section times branching ratio for generic resonances, as well as for a range of theories beyond the Standard Model.
For each test contribution from a signal model,  a simultaneous likelihood fit of data using the signal contribution plus the background function is performed,
with an additional parameter describing  the normalisation of the signal template.
The free parameters of the fit function are considered as nuisance parameters. Systematic uncertainties are also included as nuisance parameters.
To make sure the convergence of the likelihood fits, the initial parameters of the background function are set to the values
determined during the search phase, but the parameters were not fixed or constrained during the fits.
For a given mass, a number of such fits are performed for a range of possible signal yields.
The resulting likelihood function is multiplied by a flat prior to give a posterior probability density.
The 95\% quantile in the signal contribution is calculated for each pseudo-experiment determined by integration of the posterior
probability distribution; this is taken as the upper limit on the number of possible signal events in the data,
for each resonance mass and width hypothesis. This value, divided by the integrated luminosity,
provides a measure of the upper limit on the cross-section times acceptance times efficiency times branching ratio for a resonance with that mass and width.
This method was also used to determine the expected limits, along with the corresponding 1 and 2 standard deviation ($\sigma$) uncertainty bands.
The expected limits are evaluated by replacing actual data by pseudo-data generated using the background  function determined  in the search phase.
This limit-setting technique is described in \cite{PhysRevLett.105.161801} and was used in the previous ATLAS papers \cite{Aad:2011fq,Aad:2011aj,ATLAS:2012pu,Aad:2014aqa,highmass,Aaboud:2017yvp,Aaboud:2018fzt,Aad:2019hjw,Aaboud:2018fzt,Aaboud:2019zxd}.

Detector-level limits for the BSM models are corrected by the acceptance and efficiency as a function of the mass.
The acceptance is defined by the $\pt$ and $\eta$ requirements on the leptons and jets, and the requirement
on the minimum dijet invariant mass used in this analysis.
The acceptance is typically 40\% for the lowest mass point, and increases to 60--90\% for the highest mass,
depending on the model.
The efficiency correction includes various instrumental effects, such as the efficiencies for the trigger,
lepton identification and lepton reconstruction efficiencies.
A typical efficiency, averaged over the two lepton flavours, is 65--75\%,
depending on the particle mass and the type of BSM model.
The efficiency is somewhat lower (about 50\%) for the $tbH^+$ channel than for the other models
due to a more complex final state leading to fewer isolated leptons.
 
\section{Systematic uncertainties}
\label{sec:sys}
 
The systematic uncertainties considered include those associated with the background description,
the jet energy scale, jet energy resolution, lepton reconstruction, and luminosity.

The effects of jet energy scale (JES) and jet energy resolution (JER)
uncertainties~\cite{Aaboud:2017jcu} are estimated using signal model MC events.
These uncertainties cause shifts in the dijet masses by as much as $\pm 1.4\%$.
The combined effect from all systematic uncertainties leads
to a 6\% increase in the limits relative to the limits without uncertainties. This
increase  is typically within the $1\sigma$ band around the expected limits shown later.
The uncertainty associated with the JES and JER dominates the total systematic uncertainty for $\mjj\lesssim 2$~TeV.
At $\mjj>2$~TeV the uncertainties associated with the background description become comparable.
The effect of lepton energy scale uncertainties is found to be negligible.
Systematic effects from the lepton trigger, and from lepton identification and reconstruction are taken into
account by assigning a constant systematic uncertainty of 1\%.
This group of uncertainties accounts for differences between data and MC modelling.
The effect of the trigger on the shape of the $\mjj$ distribution is found to be negligible.
An additional uncertainty in the limits is associated with variations of the shapes
of the $\mjj$ signal distributions due to the PDF choice \cite{Aaboud:2017yvp}.
This arises from the fact that PDF uncertainty affects the angular distributions between the two jets, thus
affecting simulated signal shapes used to derive the limits.
Such shape-related systematic
effects on the $\mjj$ distribution are accounted for by assigning a 1\% PDF uncertainty to the calculated limits~\cite{Aaboud:2017yvp}.
The PDF uncertainty is found to have a negligible effect on the acceptance corrections, which are dominated by the selection cut on the leptons.
The uncertainty of 1.7\% in the integrated luminosity is also accounted for.
 
For the generic Gaussian signals, JES and JER systematic uncertainties from the $W'\to W Z' \to \ell\nu q\bar{q}$ MC
events are used, parameterised as a function of mass to generate
uncertainties for masses that are not covered by the $W'\to W Z' \to l\nu q\bar{q}$ samples.
As a cross-check, other models predicting different signal widths are also used, but no significant
differences are observed.
 
The uncertainties arising from imperfect knowledge of the background shapes are
estimated by using an alternative fit function, given by
Eq.~(\ref{eq:function}), with the replacement
$p_{5}\ln^{2}x \rightarrow p_5/\sqrt{x}$.
In the region $\mjj < 2$~TeV, the uncertainty related to this alternative
function is negligible compared to the statistical variation of the nominal fit.
The background shape uncertainty becomes comparable to the size of the
statistical variations of the fit function at higher values of $\mjj$.
To account for the effects of the fit function choice on the extracted limits, the largest difference in the event
yields  between the nominal and the alternative background hypothesis is taken
as a systematic uncertainty.
The possibility of statistical biases related to the choice of functional form
used to estimate the background is investigated using a closure test.
To do this, pseudo-random distributions were created according to
Eq.~(\ref{eq:function}) with the parameters obtained from the data and allowing
bin-by-bin fluctuations.
These were fitted using  Eq.~(\ref{eq:function}) plus a Gaussian (signal)
component with various fixed mean and width values, similar to what is done in
the limit calculations.
The distributions of the amplitude of the Gaussian components in these fits have
mean values close to zero for all masses and widths, with RMS values indicating
a negligible effect on the limits.
This was also verified using more complex signal shapes predicted by the $H^+$ model.

Several theoretical uncertainties  associated with the $H^+$ model are considered.
The uncertainty due to the choice of PDF is found to have a negligible effect on the $\mjj$ distribution for $H^+$ masses below $2$~TeV used
for the calculation of limits in this analysis.
The narrow-width approximation used by  \MGMCatNLO for $\tan \beta=0.5$ also has a
negligible impact on the limit presented in this paper.
This is checked by applying a smearing of the $\mjj$ distribution using the Breit--Wigner distribution with a  width
of $18\%$ as predicted in the 2HDM model.  The polarisation effect for top-quark production was studied using leading-order QCD simulations
since no spin dependence is implemented in \MGMCatNLO at NLO.
No statistically observable effect on the $\mjj$ distribution is found.

 
\section{Results}
\label{sec:result}
 
\begin{figure}
\begin{center}
\includegraphics[width=0.8\textwidth]{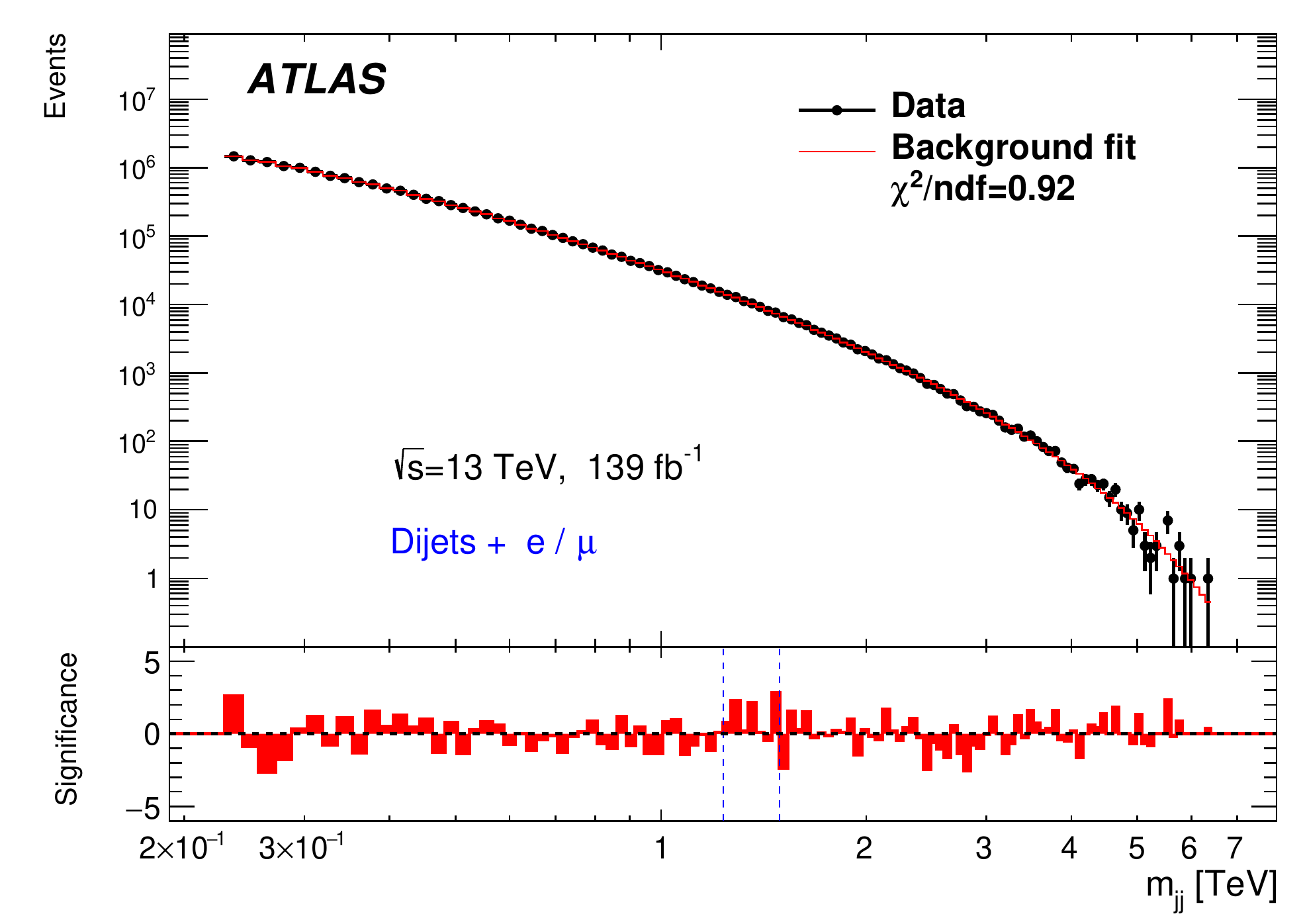}
\end{center}
\caption{Dijet invariant-mass distribution from the 2015--2018 data, from events with a high-$\pt$
lepton ($e$+$\mu$ combined).
The distribution is calculated from the two leading jets selected from events
with at least one isolated lepton with $p_\text{T}^\ell>60$~GeV.
Also shown is the result of the fit with the five-parameter background function.
The lower panel shows the bin-by-bin significances of deviations from the background hypothesis.
The largest deviation reported by BumpHunter  is indicated by the
vertical dashed lines.
The global $p$-value of this deviation is 0.31.
}
\label{fig:pub_mjj}
\end{figure}
 
\begin{figure}
\begin{center}
 
\subfloat[] {
\includegraphics[width=0.8\textwidth]{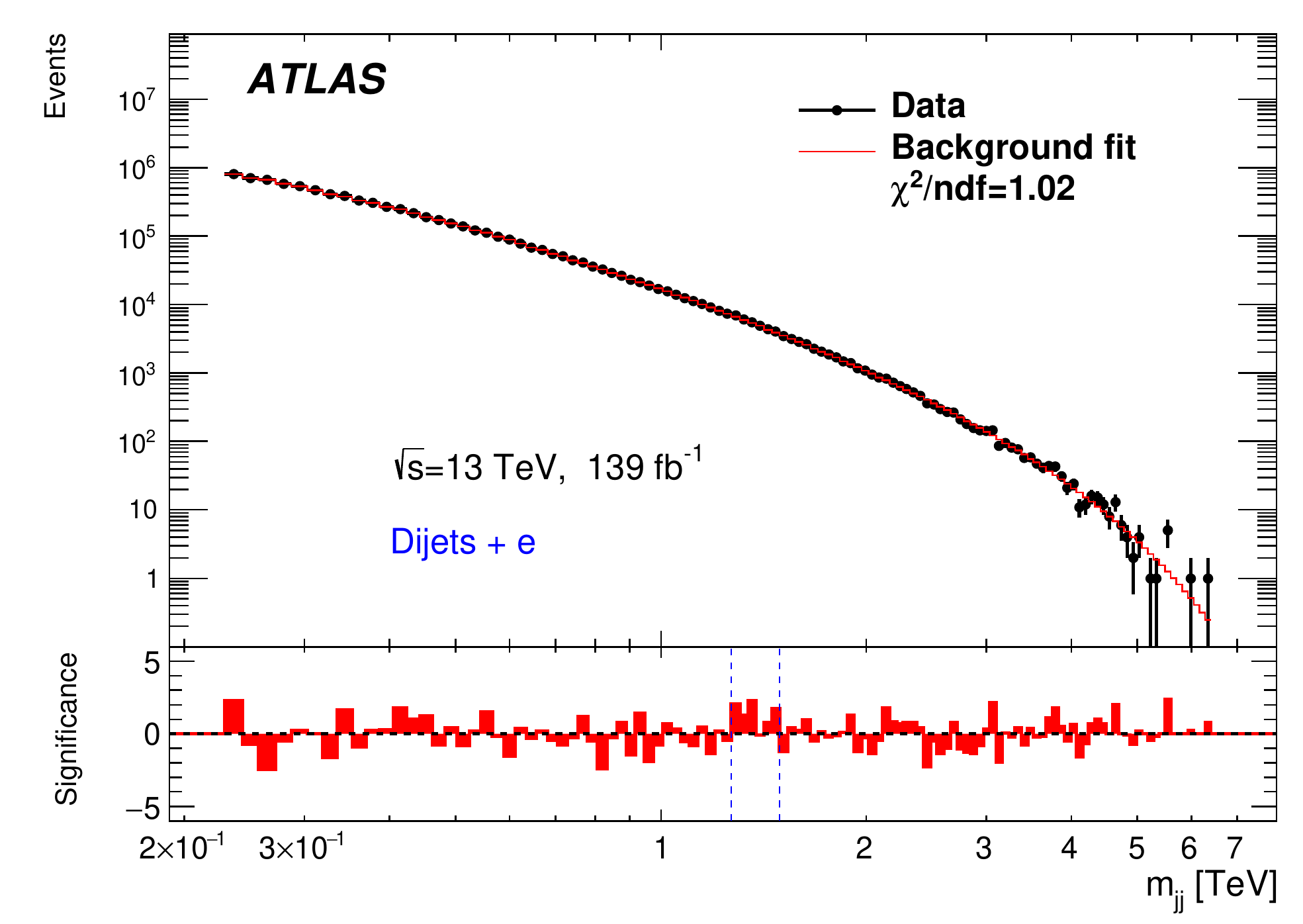}\hfill
}
 
\subfloat[] {
\includegraphics[width=0.8\textwidth]{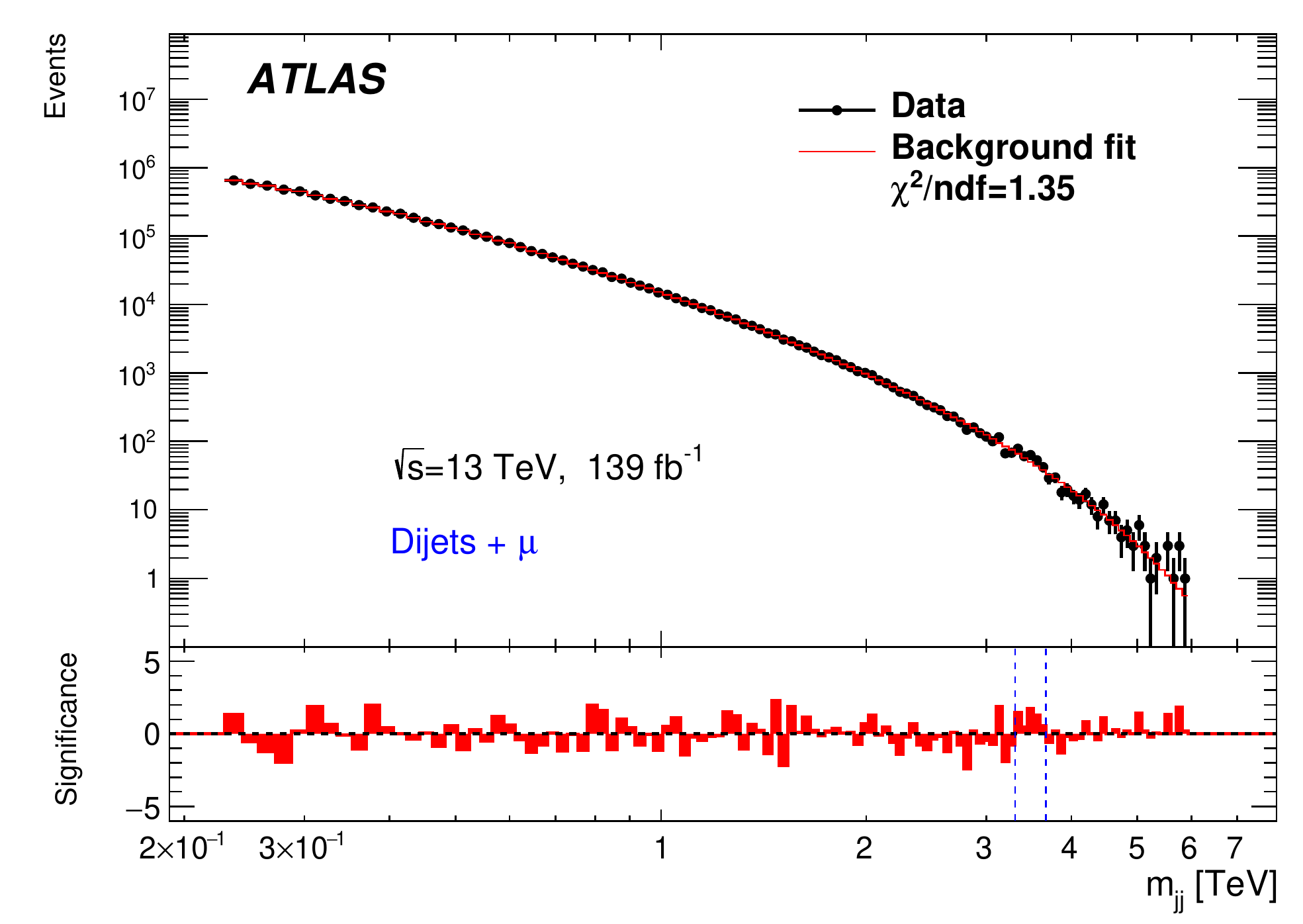}
}
\end{center}
\caption{Dijet invariant-mass distributions for events with a high-$\pt$ (a) electron or (b) muon.
The distributions are calculated from the two leading jets selected from events
with at least one isolated lepton with $p_\text{T}^\ell>60$~GeV.
In each case, the result of the fit to the five-parameter background function is also shown.
The lower panels show the bin-by-bin significances of deviations from the background hypothesis.
The largest deviations reported by BumpHunter are indicated by the
vertical dashed lines.
The global $p$-values of the deviations are 0.12 (for electrons) and 0.63 (for muons).
}
\label{fig:pub_mjj_separate}
\end{figure}

Figure~\ref{fig:pub_mjj} shows the $\mjj$ distribution obtained from the selected events
in the combined electron-plus-muon channel.
The result of applying the BumpHunter procedure
is also shown, using a background fit with the five-parameter function from Eq.~(\ref{eq:function}).
The data are well described by the fit function, with $\chi^2=99.8$ and the number of degrees
of freedom (ndf) of 109 leading to $\chi^2/\mathrm{ndf}=0.92$.
The lower panel shows the significances \cite{Choudalakis_2012} of deviations from the background hypothesis, which can be approximated
by $(d_i - f_i)/\delta$, where $d_i$ is
the value of the data points, $f_i$ is the fit value, and $\delta$ is an uncertainty.
This uncertainty includes statistical and systematic uncertainties
of the data points and the value of the fit in the $i$th bin.
These significances are consistent with a normal  distribution
with a mean of zero and unit width (not shown).
Figure~\ref{fig:pub_mjj_separate} shows the results of the analysis applied separately to events
containing a high-$\pt$ (a) electron or (b) muon, including the results from BumpHunter.
The $\chi^2/\mathrm{ndf}$ values are indicated on each figure.
 
The largest deviation of the data from the background-only hypothesis reported by BumpHunter in the combined channel is near 1.3~TeV,
with a local $p$-value of $10^{-3}$, corresponding to a significance of $2.8$ standard deviations.
The second largest deviation
near 400~GeV has a local significance of $1.3\sigma$.
Accounting for the look-elsewhere effect, the global $p$-value for the  largest deviation for the electron-plus-muon channel
is $0.3$, leading to a significance of
$0.5\sigma$.
This deviation from the background hypothesis is consistent with a statistical fluctuation,
and  results mainly from the electron channel.
The largest deviation from the background hypothesis for the muon channel shown in Figure~\ref{fig:pub_mjj_separate}(b)
is near 3.5~TeV, with the global $p$-value of 0.6.
 
In the absence of any significant signals indicating the presence of new phenomena beyond the SM,
limits are set in the manner described in Section~\ref{sec:ana}.
The limits include the systematic uncertainties described in Section~\ref{sec:sys}.
 
\begin{figure}[t!]
\begin{center}
\includegraphics[width=0.8\textwidth]{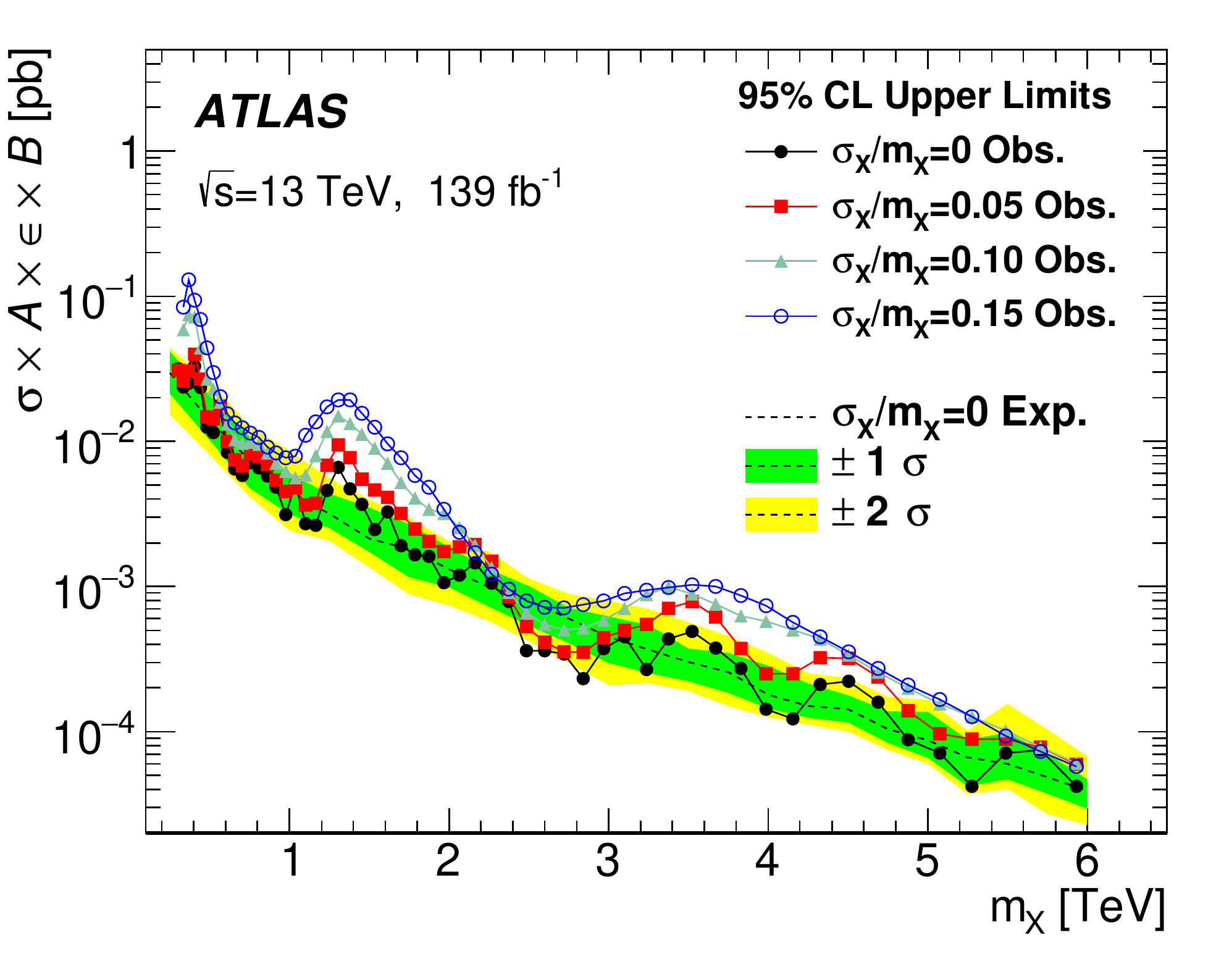}\hfill
\end{center}
\caption{
The $95\%$ CL observed limits for
a hypothetical particle $\rm{X}$ resulting in a contribution to the observed
$\mjj$ distribution with a Gaussian shape and various widths $\sigmaX$.
The $\mjj$ distribution is obtained from the two leading jets in events with at
least one isolated lepton with $p_\text{T}^\ell>60$~GeV.
The limits, presented for the fine steps in masses that correspond to the bin sizes times two,
are calculated assuming widths of the Gaussian signal corresponding to 0\%, 5\%, 10\% and  15\%
of the signal mass.
For the latter two cases,
the points below $\mjj=0.3$~TeV are excluded since
the signal would, in part, be at masses below the minimum value considered here.
The limits are set on  the cross-section times the acceptance $A$, the efficiency $\epsilon$ and
branching ratio $B$.
The expected limit and the corresponding
$\pm 1\sigma$ and $\pm 2\sigma$ bands are shown for the $\sigmaX/\mX$=0 signals.
}
\label{fig:pub:gaussians_all}
\end{figure}

Figure~\ref{fig:pub:gaussians_all} shows the $95\%$ CL observed limits for
a hypothetical particle $\rm{X}$ resulting in a contribution to the observed
$\mjj$ distribution with a Gaussian shape and various widths $\sigmaX$.
The limits are presented as a function of the mass $\mX$ in steps that
correspond to the bin sizes times two.
The background hypothesis is defined as the five-parameter fit described earlier.
The expected limit and the corresponding
$\pm 1\sigma$ and $\pm 2\sigma$ bands are shown for $\sigmaX /\mX$=0 signals.
Appendix~\ref{appendix:broad} shows
the expected limit and the corresponding
$\pm 1\sigma$ and $\pm 2\sigma$ bands for the $\sigmaX/\mX$=0.15 signals.
Contributions from a Gaussian-shaped signal with minimum effective cross-sections ranging from
approximately 100~fb to 0.1~fb are excluded in the mass range of 0.25--6~\TeV.
 
The oscillations in the observed limits shown in Figure~\ref{fig:pub:gaussians_all}
for wide Gaussian signals are due to correlations between the points in the limit calculation, which uses the background
function with unconstrained parameters. The correlation length between neighbouring mass points is proportional
to the width of the resonances  used for the limits. The correlation lengths increase with the width of the assumed Gaussian signal.
 
As a check, the statistical significance of the largest excess is studied under the signal-plus-background hypothesis.
The function used to describe the data is constructed by adding a Gaussian distribution to the background shape
described by Eq.~(\ref{eq:function}).
All the parameters but the Gaussian width are allowed to vary during the minimisation procedure.
The local significance of the largest excess, calculated from the Gaussian normalisation factor and its uncertainty,
does not exceed $2.6\sigma$, i.e. it is smaller than the local significance reported by the BumpHunter.
The same conclusion is obtained by using the test based on the likelihood ratio
$-2 \ln L(0)/L(1)$, with $L(0)$ and $L(1)$ being the likelihoods
of the null hypothesis and the signal-plus-background hypothesis from the global
likelihood fits of the data,  assuming the asymptotic approximation \cite{Cowan:2010js}.
Systematic uncertainties are not included in this check.
 
\subsection{Limits on BSM models}

Exclusion limits are also set for the four BSM models discussed earlier.
As discussed before, the analysis requires well-reconstructed jets and
leptons, without additional selections tuned to the specific model under
study.
For the limit calculations, the shape of the $\mjj$
distribution for each model considered is taken from MC simulation, after detector simulation
and reconstruction.
The $\mjj$ distributions for the $\pi_T$ and $Z'$ decays have Gaussian-like shapes around the nominal
generated masses, with a HWHM ranging from 10\% to 20\% in their Gaussian cores.
The $H^+$ model has an $\mjj$ shape with a HWHM of about $30\%$, with the peak position
shifted to a lower mass compared to $m_{H^+}$ as discussed in Section~\ref{sec:mc}.
 
The Bayesian limits are calculated using the $\mjj$ distribution with the background description
discussed earlier, and then are corrected by the acceptance and efficiency corrections.
The values of the calculated limits depend on the shape of the signal $\mjj$ distributions, e.g.
broader distributions typically lead to higher values of the  limits.
All calculated limits account for the previously discussed systematic uncertainties.
In each case, the minimum $\mjj$ value for the limits is selected to ensure that the signal acceptance
is at least $30\%$ for the mass points considered.

\begin{figure}
\begin{center}
\subfloat[] {
\includegraphics[width=0.46\textwidth]{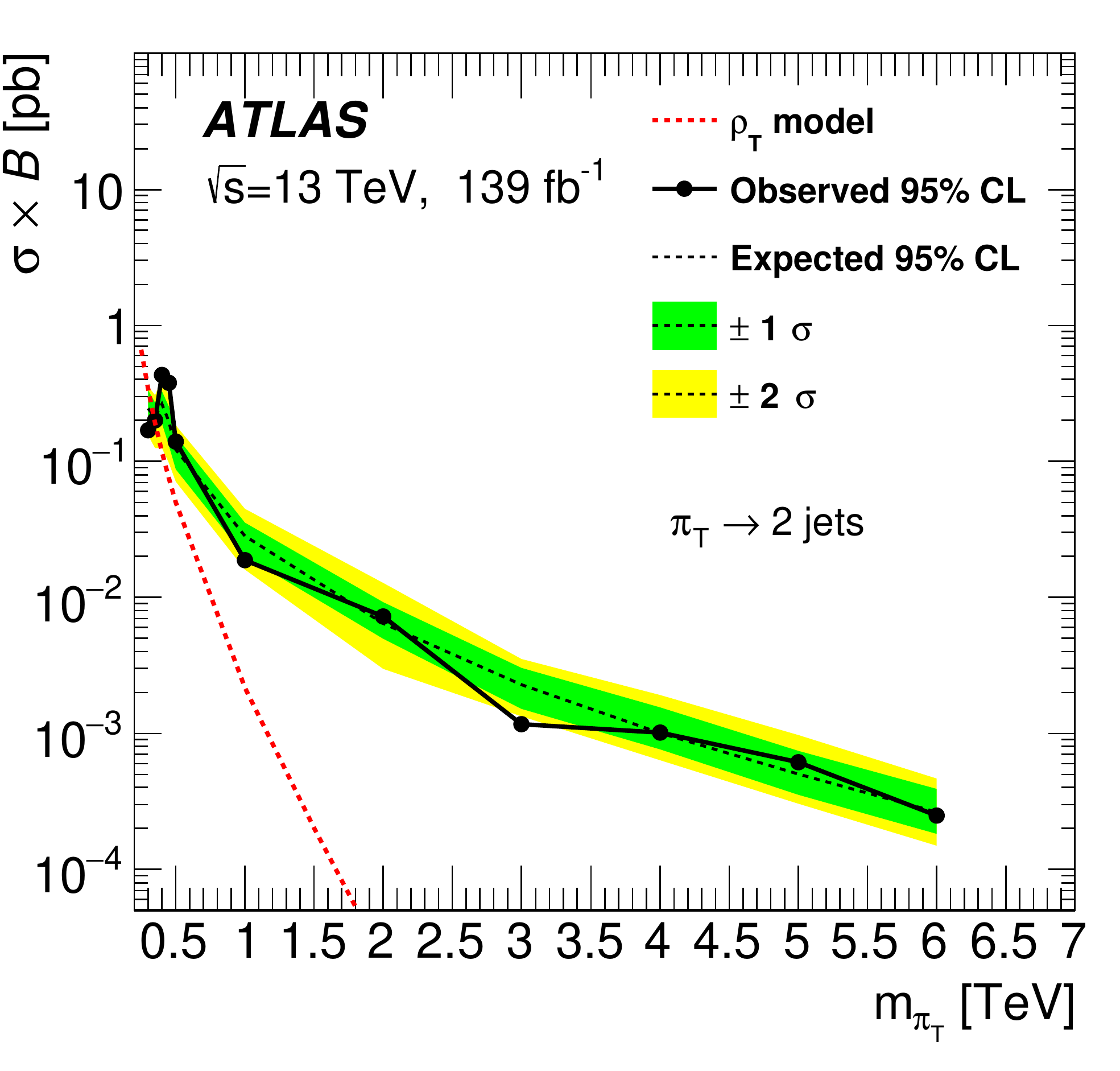}\hfill
}
\subfloat[] {
\includegraphics[width=0.46\textwidth]{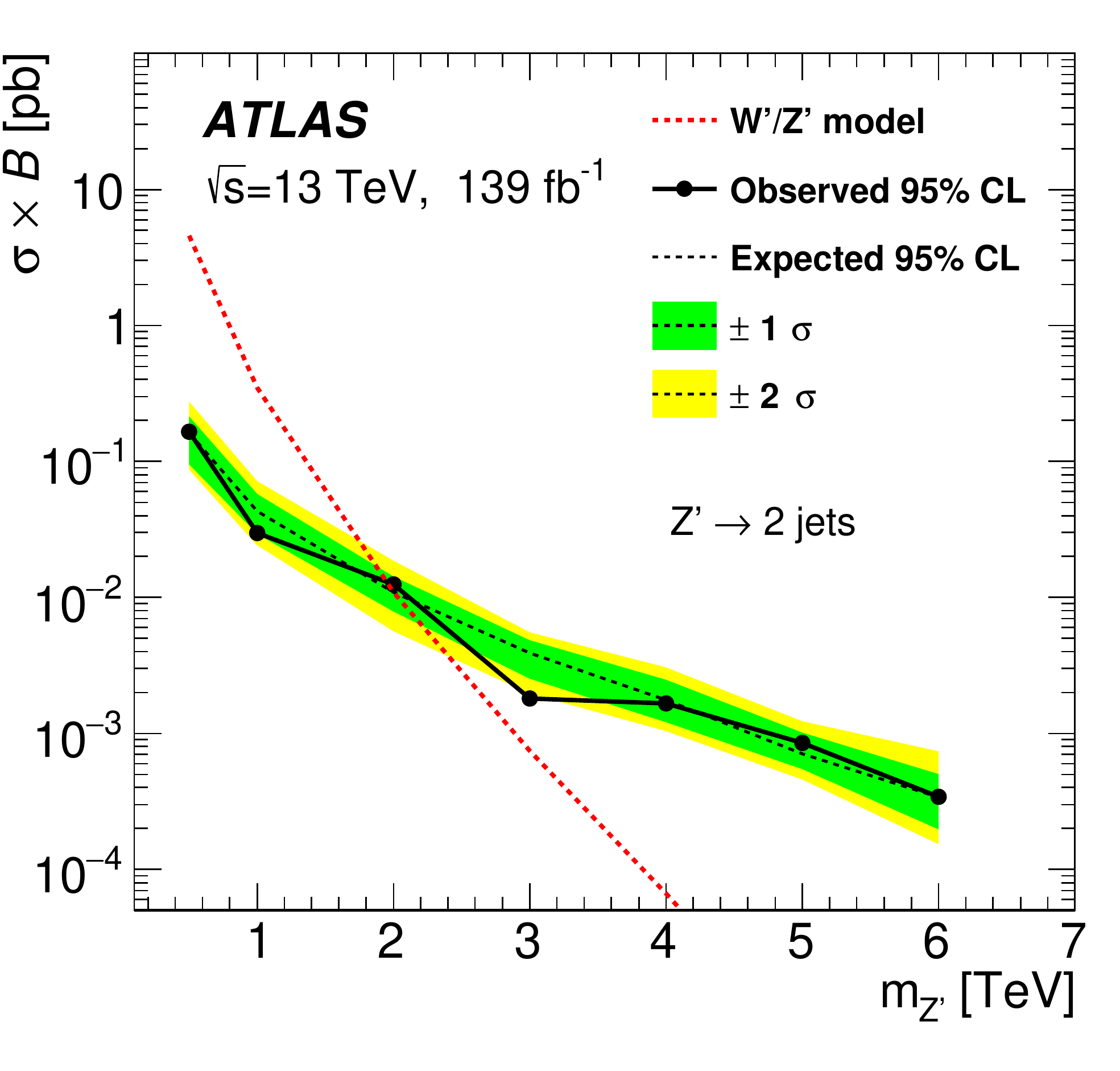}\hfill
}
 
\subfloat[] {
\includegraphics[width=0.46\textwidth]{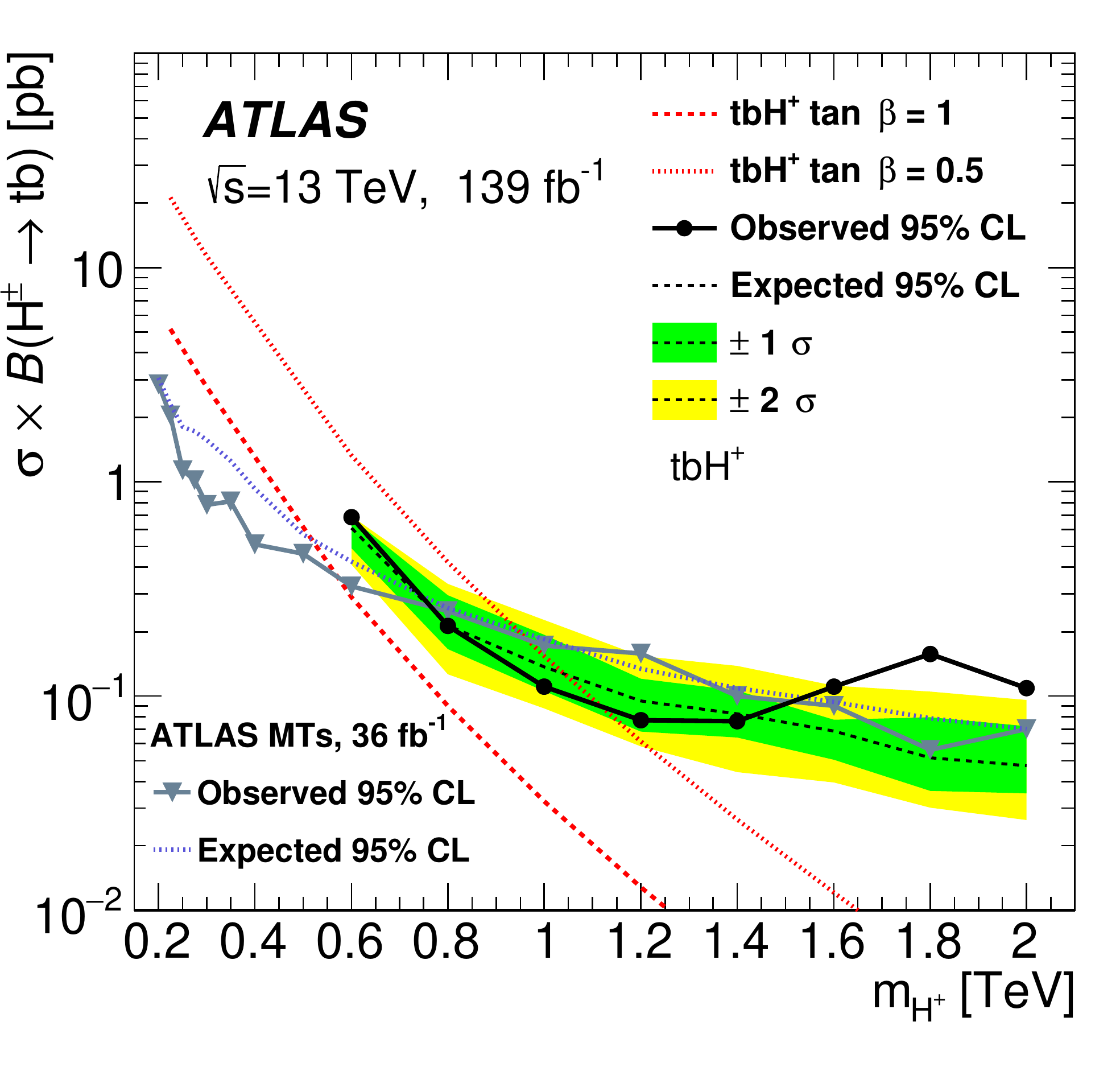}\hfill
}
\subfloat[] {
\includegraphics[width=0.46\textwidth]{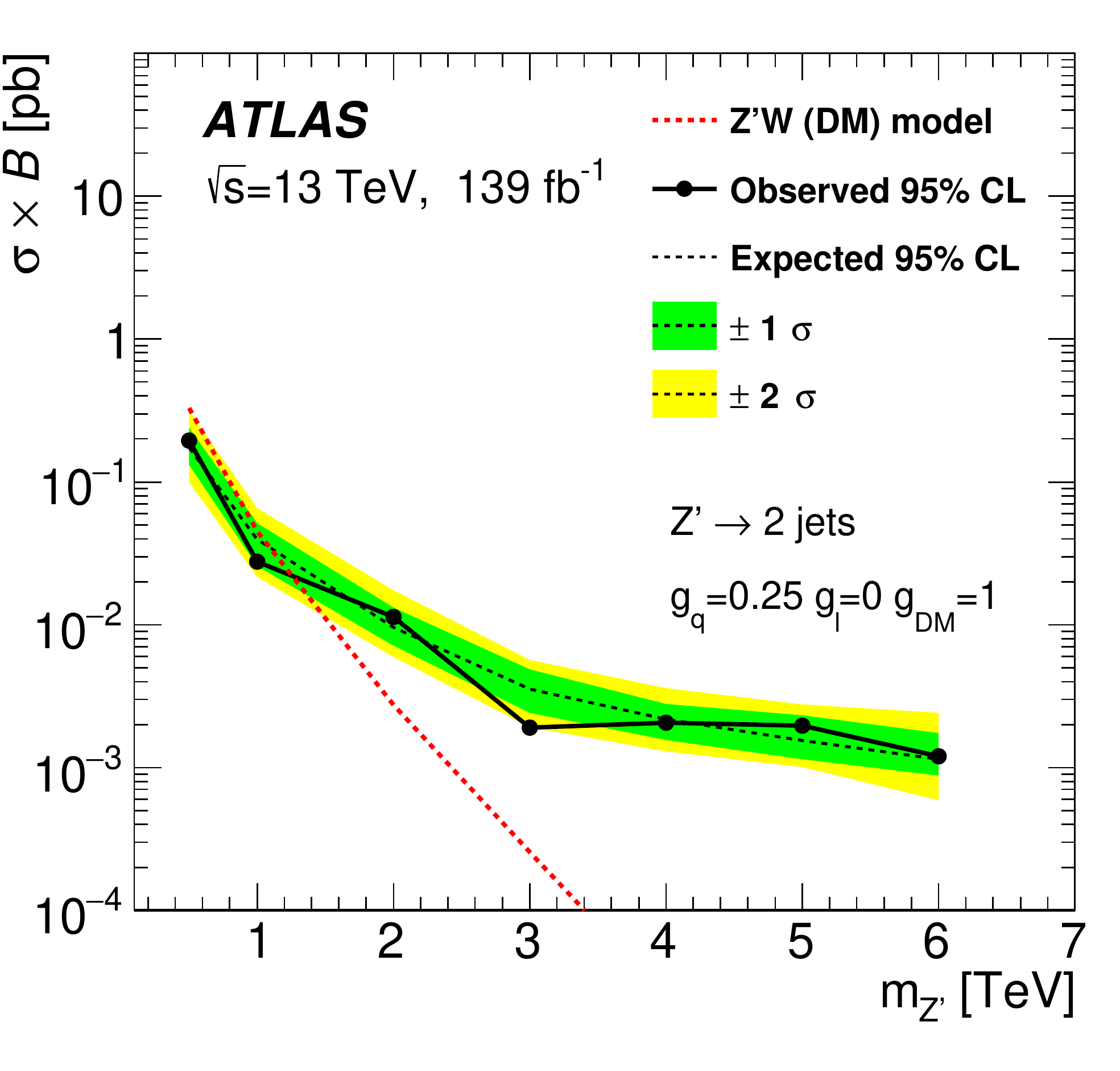}
}
 
\end{center}
\caption{
Observed (filled circles) and expected (dotted line with uncertainty bands)
95\% credibility-level upper limits on the cross-section ($\sigma$)
times branching ratio ($B$) for
(a) the techicolor model with
production of $\rho_T$ decaying into $\pi_T W^{\pm}$, (b)
$W' \to Z' W^{\pm}$ production
in the Sequential Standard Model, (c) the $tbH^+$  model
for $\tan\beta=1$ (thick  red dashed line) and $\tan\beta=0.5$ (thin red dashed line),
(d) the simplified dark-matter model.
Figure (c) also shows the expected and observed limits (without indicating the 1 and $2\sigma$ bands) from the early Run~2 paper~\cite{Aaboud:2018cwk} based on
the multivariate techniques (MTs) in the signal regions to enhance the separation of signal from background.
The presented limits are obtained using
two leading jets in events with at least one isolated lepton with $p_\text{T}^\ell>60$~GeV.
}
\label{fig:limit:final}
\end{figure}

The limits on the $\rho_T \to \pi_T W^{\pm}$ and $W' \to Z' W^{\pm}$ signal models are
shown in Figure~\ref{fig:limit:final}(a) and Figure~\ref{fig:limit:final}(b), respectively.
For the technicolor model, in which the mass of the $\rho_T$ is twice that of
the $\pi_T$, values of $m_{\pi_T}$ below 350~GeV are
excluded for the range of technipion masses considered.
A small deviation of the observed limits from the expected limits near 400~GeV is consistent with a statistical fluctuation,
as follows from the BumpHunter background-only hypothesis (see Section~\ref{sec:result}).
It is not observed for the Gaussian limits due to the minimum $\mjj$ requirement for the Gaussian signals with widths above 10\%.
The $W' \to Z' W^{\pm}$ model is excluded for $Z'$ masses up to 2~TeV, assuming the
maximal cross-section, which occurs when the mass difference between the $W'$ and $Z'$
is 250~GeV (see Section~\ref{sec:mc}).
 
The $\mjj$ distributions obtained using model-independent selection
criteria can also be used to set limits on complex decay topologies.
This is illustrated in the context of the $H^+$ signal model discussed in Section~\ref{sec:mc}.
The limits obtained
without a channel-specific selection
of this process are shown in Figure~\ref{fig:limit:final}(c).
The data exclude a charged Higgs particle
with a mass below 1.2~TeV, assuming $\tan\beta=0.5$, the 2HDM type-2 model with the four-flavour scheme
and the narrow-width approximation.
The $H^+$ limits are compared with the result~\cite{Aaboud:2018cwk} based
on multivariate techniques and an early subset of the Run~2
data corresponding to 36.1~fb$^{-1}$ of integrated luminosity.
On average, the observed and expected limits in the mass range 0.8~TeV--1.4~TeV are about a factor of two better, and
the highest excluded $H^+$ mass is 200~GeV higher for $\tan \beta=0.5$, than for the earlier Run~2 analysis.
The observed differences in the limits between the previous analysis and the current analysis are due to differences
in experimental methods adopted in these two studies, as well as due to differences in the integrated luminosities.
The excess above the expected limit near an $H^+$ mass of 1.8~TeV
corresponds to the excess near 1.3~TeV for the background-only hypothesis, which is also observed in the Gaussian limits
as discussed in Section~\ref{sec:result}.
The limits shown in Figure~\ref{fig:limit:final}(c) are correlated
since the reconstructed width of the signal is larger than the mass difference between the limit points.
To verify the local statistical significance for the $H^+$ signal model
under the signal-plus-background hypothesis, the likelihood fit of the data with
the test statistic is performed using the asymptotic approximation~\cite{Cowan:2010js}.
Alternatively,
the significance is calculated from the amplitude of the signal component of the
signal-plus-background fit, after modelling the shape of the $\mjj$ distribution analytically.
No systematic uncertainties are included.
In all cases the local significance does not exceed $2.4\sigma$, i.e. it is smaller than the significance
reported by the background-only hypothesis discussed  earlier.

\begin{samepage}
The limits on the simplified dark-matter model, for the leptophobic couplings $g_q = 0.25$, $g_\ell=0$ and $g_\text{DM}=1$,
are shown in Figure~\ref{fig:limit:final}(d).
These exclude $Z'$ masses for this model below 1.2~TeV, complementing
exclusions set previously by the ATLAS inclusive dijet search~\cite{Aad:2019hjw}
for $m_{Z'} < 1.5$~TeV.
It was checked that changing $g_\ell=0$ to a small value  ($g_\ell=0.01$) leads to a negligible effect on the presented limits.
\end{samepage}

\FloatBarrier
\section{Conclusion}
A search for resonances in dijet invariant-mass distributions is presented,
based on the analysis of events in which the jets are accompanied by at least
one isolated high-$p_\text{T}$ lepton ($e$ or $\mu$).
Events are selected from a data sample corresponding to an integrated luminosity
of \lumi~fb$^{-1}$ of proton--proton collisions at $\sqrt{s}=13$~TeV, recorded
by the ATLAS detector during Run 2 of the LHC.
 
In the dijet invariant-mass range considered, 0.22--6.3~TeV, the most
significant deviation from data-derived estimate of the Standard Model
background in the combined electron and muon channel is observed around
$\mjj=1.3$~TeV.
Taking into account both the systematic uncertainties and the look-elsewhere
effect, this excess has a $p$-value of 0.3.
The data are thus consistent with the background-only hypothesis.

This analysis has set 95\% credibility-level
upper limits on the signal cross-section times acceptance times efficiency times
branching ratio for new processes that can produce a Gaussian contribution to
the dijet invariant-mass distribution in events having at least one isolated
lepton with $p_\text{T}^\ell>60$~GeV.
Limits are calculated in different scenarios for the width of these Gaussian
signals, ranging from that determined by the detector resolution up to 15\% of
the resonance mass.
The limits obtained range from $100$~fb to $0.1$~fb for resonance masses between
$0.25$ and $6$~TeV.
 
Model-dependent limits are also set on a variety of BSM models, without the use
of additional selection criteria tailored to the specific final states
investigated.
These results exclude contributions from the $W'\to W Z'$ process in the
Sequential Standard Model for masses of the $Z'$ (decaying into jet pairs) below
2~TeV, assuming the mass difference between the $W'$ and $Z'$ is 250~GeV to
maximise the cross-section for this process.
For a technicolor model in which the relationship between the $\rho_T$ and
$\pi_T$ masses maximises the cross-section for the final state of interest,
technipion masses below 350~GeV are excluded for the range of technipion masses
considered.
 
The model-dependent limits obtained without optimisation for the specific signal
models are shown to have the potential to exclude heavy states with complex
decays, such as in charged Higgs boson production in association with a top
quark, $tbH^+$.
For this model, the data exclude $H^+$ masses below 1.2~TeV assuming
$\tan\beta=0.5$.
These results complement those from the dedicated $H^+$
studies~\cite{Aaboud:2018cwk} that employ a selection optimised for the
charged Higgs event topology.
 
The data also exclude $Z'$ mediator with masses below 1.2~TeV in a simplified
Dark Matter model with leptophobic couplings
($g_q = 0.25$, $g_\ell=0$ and $g_\text{DM}=1$), in which the lepton originates
from the decay of an associated $W$ boson.
 
 
\section*{Acknowledgements}
 

We thank CERN for the very successful operation of the LHC, as well as the
support staff from our institutions without whom ATLAS could not be
operated efficiently.
 
We acknowledge the support of ANPCyT, Argentina; YerPhI, Armenia; ARC, Australia; BMWFW and FWF, Austria; ANAS, Azerbaijan; SSTC, Belarus; CNPq and FAPESP, Brazil; NSERC, NRC and CFI, Canada; CERN; CONICYT, Chile; CAS, MOST and NSFC, China; COLCIENCIAS, Colombia; MSMT CR, MPO CR and VSC CR, Czech Republic; DNRF and DNSRC, Denmark; IN2P3-CNRS and CEA-DRF/IRFU, France; SRNSFG, Georgia; BMBF, HGF and MPG, Germany; GSRT, Greece; RGC and Hong Kong SAR, China; ISF and Benoziyo Center, Israel; INFN, Italy; MEXT and JSPS, Japan; CNRST, Morocco; NWO, Netherlands; RCN, Norway; MNiSW and NCN, Poland; FCT, Portugal; MNE/IFA, Romania; MES of Russia and NRC KI, Russia Federation; JINR; MESTD, Serbia; MSSR, Slovakia; ARRS and MIZ\v{S}, Slovenia; DST/NRF, South Africa; MINECO, Spain; SRC and Wallenberg Foundation, Sweden; SERI, SNSF and Cantons of Bern and Geneva, Switzerland; MOST, Taiwan; TAEK, Turkey; STFC, United Kingdom; DOE and NSF, United States of America. In addition, individual groups and members have received support from BCKDF, CANARIE, Compute Canada and CRC, Canada; ERC, ERDF, Horizon 2020, Marie Sk{\l}odowska-Curie Actions and COST, European Union; Investissements d'Avenir Labex, Investissements d'Avenir Idex and ANR, France; DFG and AvH Foundation, Germany; Herakleitos, Thales and Aristeia programmes co-financed by EU-ESF and the Greek NSRF, Greece; BSF-NSF and GIF, Israel; CERCA Programme Generalitat de Catalunya and PROMETEO Programme Generalitat Valenciana, Spain; G\"{o}ran Gustafssons Stiftelse, Sweden; The Royal Society and Leverhulme Trust, United Kingdom.
 
The crucial computing support from all WLCG partners is acknowledged gratefully, in particular from CERN, the ATLAS Tier-1 facilities at TRIUMF (Canada), NDGF (Denmark, Norway, Sweden), CC-IN2P3 (France), KIT/GridKA (Germany), INFN-CNAF (Italy), NL-T1 (Netherlands), PIC (Spain), ASGC (Taiwan), RAL (UK) and BNL (USA), the Tier-2 facilities worldwide and large non-WLCG resource providers. Major contributors of computing resources are listed in Ref.~\cite{ATL-SOFT-PUB-2020-001}.
 
We thank Dr. G.~Bodwin for the studies of the analytic properties of Eq.~(\ref{eq:function}).

\clearpage
\appendix
\part*{Appendix}
\addcontentsline{toc}{part}{Appendix}
\section{Dijet invariant mass in the LE-CR region}
\label{appendix:le}
 
\begin{figure}[h]
\begin{center}
\includegraphics[width=0.8\textwidth]{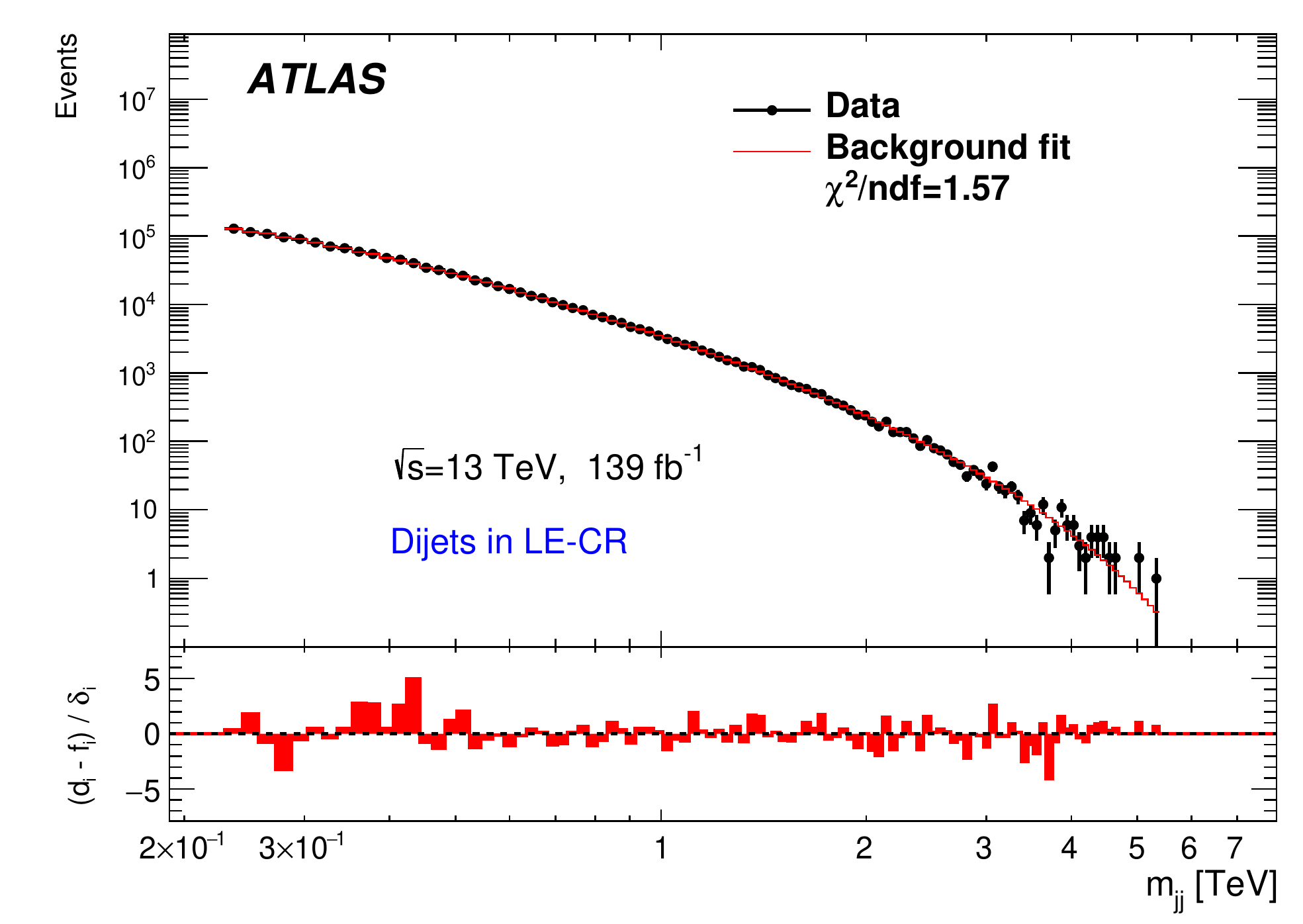}
\end{center}
\caption{Dijet invariant-mass distribution for the LE-CR region for the 2015--2018 data.
Also shown is the result of the fit with the five-parameter background function.
The lower panel shows the fit residuals divided by the uncertainty on data points. No systematic uncertainties are included.
}
\label{fig:pub_mjj_le}
\end{figure}

Figure~\ref{fig:pub_mjj_le} illustrates the distribution of events
in the LE-CR defined
for the data. Also shown is the five-parameter fit function used to describe this region.
 
\newpage
\section{Expected limits for broad signals}
\label{appendix:broad}

\begin{figure}[h]
\begin{center}
\includegraphics[width=0.8\textwidth]{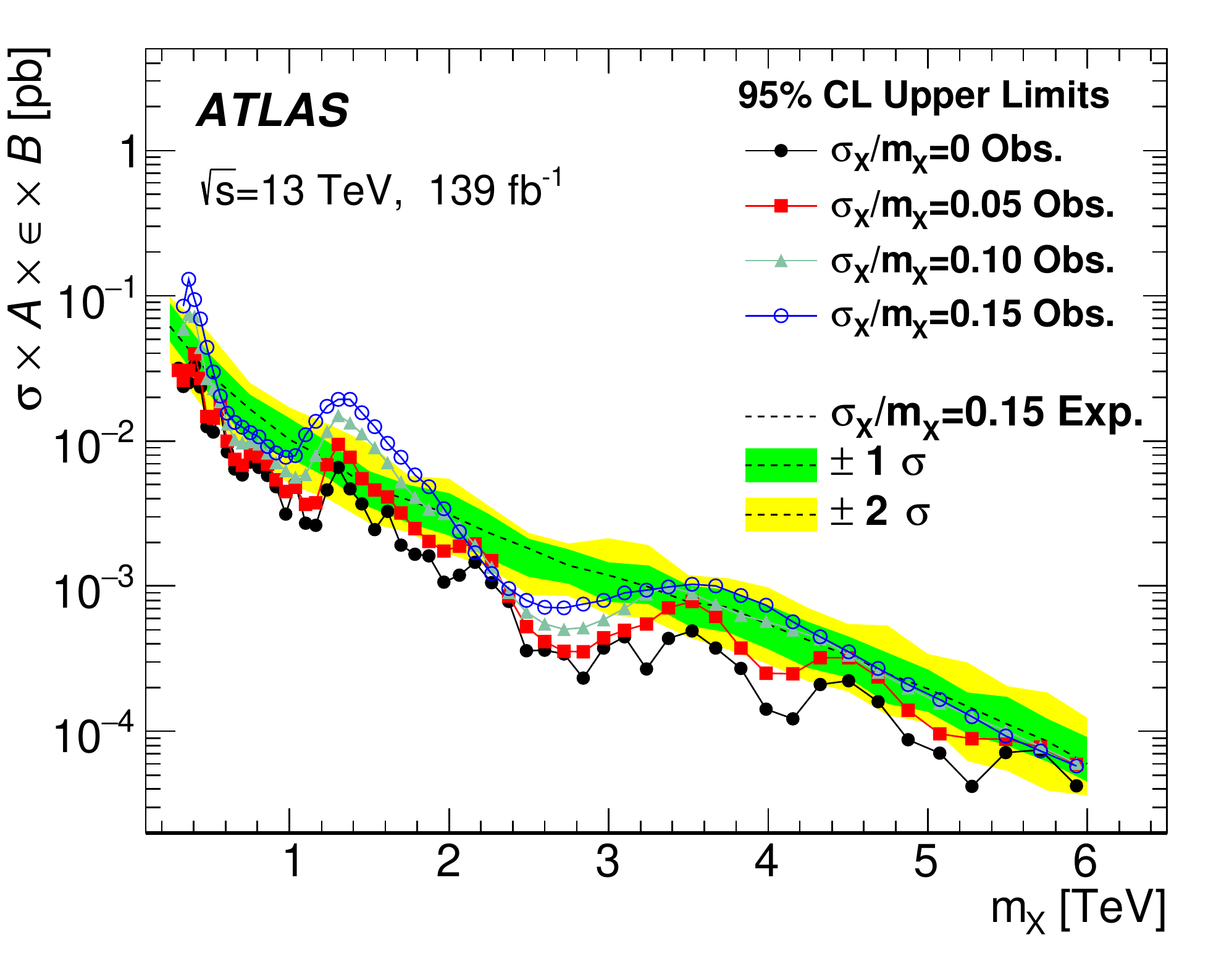}\hfill
\end{center}
\caption{
The $95\%$ CL observed limits for
a hypothetical particle $\rm{X}$ resulting in a contribution to the observed
$\mjj$ distribution with a Gaussian shape and various widths $\sigmaX$.
The $\mjj$ distribution is obtained from the two leading jets in events with at
least one isolated lepton with $p_\text{T}^\ell>60$~GeV.
The limits, presented for the fine steps in masses that correspond to the bin sizes times two,
are calculated assuming widths of the Gaussian signal corresponding to 0\%, 5\%, 10\% and  15\%
of the signal mass.
The limits are set on  the cross-section times the acceptance $A$, the efficiency $\epsilon$ and
branching ratio $B$.
The expected limit and the corresponding
$\pm 1\sigma$ and $\pm 2\sigma$ bands are shown for the $\sigmaX/\mX$=0.15 signals.
}
\label{fig:pub:gaussians_all_w15}
\end{figure}

Figure~\ref{fig:pub:gaussians_all_w15} shows the comparison
of the observed limits with the expected limit, including the corresponding
$\pm 1\sigma$ and $\pm 2\sigma$ bands, for the $\sigmaX/\mX$=0.15 signals.
The observed limits are the same as for Figure~\ref{fig:pub:gaussians_all}.

\clearpage
 
\printbibliography
 
\clearpage \input{atlas_authlist}

\end{document}

%% file: atlas_authlist.tex
 
\begin{flushleft}
\hypersetup{urlcolor=black}
{\Large The ATLAS Collaboration}

\bigskip

\AtlasOrcid[0000-0002-6665-4934]{G.~Aad}$^\textrm{\scriptsize 102}$,    
\AtlasOrcid[0000-0002-5888-2734]{B.~Abbott}$^\textrm{\scriptsize 129}$,    
\AtlasOrcid{D.C.~Abbott}$^\textrm{\scriptsize 103}$,    
\AtlasOrcid[0000-0002-2788-3822]{A.~Abed~Abud}$^\textrm{\scriptsize 36}$,    
\AtlasOrcid[0000-0002-1002-1652]{K.~Abeling}$^\textrm{\scriptsize 53}$,    
\AtlasOrcid[0000-0002-2987-4006]{D.K.~Abhayasinghe}$^\textrm{\scriptsize 94}$,    
\AtlasOrcid[0000-0002-8496-9294]{S.H.~Abidi}$^\textrm{\scriptsize 167}$,    
\AtlasOrcid[0000-0002-8279-9324]{O.S.~AbouZeid}$^\textrm{\scriptsize 40}$,    
\AtlasOrcid{N.L.~Abraham}$^\textrm{\scriptsize 156}$,    
\AtlasOrcid[0000-0001-5329-6640]{H.~Abramowicz}$^\textrm{\scriptsize 161}$
% ...
\bigskip
\\

% ...
$^{36}$CERN, Geneva; Switzerland.\\
% ...
$^{40}$Niels Bohr Institute, University of Copenhagen, Copenhagen; Denmark.\\
% ...
$^{53}$II. Physikalisches Institut, Georg-August-Universit\"{a}t G\"ottingen, G\"ottingen; Germany.\\
% ...
$^{94}$Department of Physics, Royal Holloway University of London, Egham; United Kingdom.\\
% ...
$^{102}$CPPM, Aix-Marseille Universit\'e, CNRS/IN2P3, Marseille; France.\\
$^{103}$Department of Physics, University of Massachusetts, Amherst MA; United States of America.\\
% ...
$^{129}$Homer L. Dodge Department of Physics and Astronomy, University of Oklahoma, Norman OK; United States of America.\\
% ...
$^{156}$Department of Physics and Astronomy, University of Sussex, Brighton; United Kingdom.\\
% ...
$^{161}$Raymond and Beverly Sackler School of Physics and Astronomy, Tel Aviv University, Tel Aviv; Israel.\\
% ...
$^{167}$Department of Physics, University of Toronto, Toronto ON; Canada.\\
% ...

\end{flushleft}
